\documentclass[acmsmall, screen]{acmart}
\usepackage{amsmath,amsfonts}
\usepackage{algorithmic}
\usepackage{graphicx}
\usepackage{textcomp}
\usepackage{xcolor}
\usepackage{tabularx}
\usepackage{subcaption}
\usepackage[font=small,skip=4pt]{caption}
\usepackage{enumitem}
\usepackage{siunitx}
\usepackage{stfloats}
\usepackage[ruled,vlined]{algorithm2e}
\usepackage{color, colortbl}
\usepackage{multirow}
\acmJournal{TIOT}
\acmYear{2021} \acmVolume{3} \acmNumber{1} 
\acmMonth{9} \acmPrice{15.00}\acmDOI{10.1145/3485434}
\usepackage{url}
\usepackage[normalem]{ulem}
\usepackage{amsthm}
\usepackage{gensymb}
\usepackage{pifont}

\newcounter{definition}
\newenvironment{definition}[1][]{\refstepcounter{definition}\par\medskip
\vspace{-1mm}
\noindent \textbf{Definition~\thedefinition.#1} \rmfamily}{\smallskip}

\AtBeginDocument{%
  \providecommand\BibTeX{{%
    \normalfont B\kern-0.5em{\scshape i\kern-0.25em b}\kern-0.8em\TeX}}}

\newcommand{\mgait}[1]{{\textit{MGait}}} %

\setcopyright{none}
\begin{document}

\title[\textit{MGait}: Model-Based Gait Analysis Using Wearable Bend and Inertial Sensors]{\textit{MGait}: Model-Based Gait Analysis Using Wearable Bend and Inertial Sensors}

\author{Sizhe An}
\affiliation{%
  \institution{University of Wisconsin-Madison}
  \streetaddress{1415 Engineering Dr}
  \city{Madison}
  \state{Wisconsin}
  \postcode{53706}
  \country{USA}}
\email{sizhe.an@wisc.edu}

\author{Yigit Tuncel}
\affiliation{%
  \institution{University of Wisconsin-Madison}
  \streetaddress{1415 Engineering Dr}
  \city{Madison}
  \state{Wisconsin}
  \postcode{53706}
  \country{USA}}
\email{tuncel@wisc.edu}

\author{Toygun Basaklar}
\affiliation{%
  \institution{University of Wisconsin-Madison}
  \streetaddress{1415 Engineering Dr}
  \city{Madison}
  \state{Wisconsin}
  \postcode{53706}
  \country{USA}}
\email{basaklar@wisc.edu}

\author{Gokul K. Krishnakumar}
\affiliation{%
  \institution{Arizona State University}
  \streetaddress{781 S Terrace Rd}
  \city{Tempe}
  \state{Arizona}
  \postcode{85287}
  \country{USA}}
\email{gkkrishn@asu.edu}

\author{Ganapati Bhat}
\affiliation{%
  \institution{Washington State University}
  \streetaddress{355 NE Spokane St}
  \city{Pullman}
  \state{Washington}
  \postcode{99164}
  \country{USA}}
\email{ganapati.bhat@wsu.edu}

\author{Umit Y. Ogras}
\affiliation{%
  \institution{University of Wisconsin-Madison}
  \streetaddress{1415 Engineering Dr}
  \city{Madison}
  \state{Wisconsin}
  \postcode{53706}
  \country{USA}}
\email{uogras@wisc.edu}




\begin{abstract}
%
Movement disorders, such as Parkinson's disease, affect more than 10 million people worldwide. Gait analysis is a critical step in the diagnosis and rehabilitation of these disorders. Specifically, step and stride lengths provide valuable insights into the gait quality and rehabilitation process. 
However, traditional approaches for estimating step length are not suitable for continuous daily monitoring since they rely on special mats and clinical environments. 
To address this limitation, this paper presents a novel and practical step-length estimation technique using low-power wearable bend and inertial sensors. Experimental results show that the proposed model estimates step length with 5.49\% mean absolute percentage error and provides accurate real-time feedback to the user.

\end{abstract}




\maketitle

\section{Introduction}
\vspace{-0.5mm}
\label{sec:introduction}

Movement disorders are one of the leading causes of functional disability in the elderly population. More than 10 million people worldwide suffer from movement disorders, such as Parkinson's disease~(PD); more than 900,000 patients are expected to be diagnosed with PD in the United States by the year 2020~\cite{marras2018prevalence,bhat2019openhealth}.
Gait impairment and instability are among the most common symptoms in movement disorder patients and stroke survivors~\cite{hsu2003analysis}. Therefore, gait function analysis plays an important role in treatment and rehabilitation.

Gait function analysis provides valuable insight into a patient's symptoms and rehabilitation, by evaluating metrics, such as step length, stride length, and gait velocity~\cite{pirker2017gait,anderson_shi_2019}. In particular, step length is used to analyze the symmetry of gait. 
While step length remains nearly constant for a healthy person, 
it varies as the left and right feet alternate
for a patient with gait asymmetry. 
This difference is used as an important feature to evaluate the symmetry of gait and monitor the progress of rehabilitation~\cite{pirker2017gait}. 

Several clinical studies use the GAITRite system~\cite{gaitrite}, a pressure-sensitive walking mat, to analyze the gait parameters~\cite{gonzalez_2016gaitritevisionbased, mcdonough2001validity, webster2005validity}.
While GAITRite can provide a 98\%-99\% accuracy, it cannot be used to continuously monitor the patient's gait after they leave the clinic. 
To address this limitation, recent work employs wearable sensors for gait analysis~\cite{wang_sun_li_liu_2018,anderson_shi_2019,wu_wang_pottie_2015}. Most of these studies mount a number of inertial motion units~(IMU), which typically incorporate a 3-axis accelerometer and a 3-axis gyroscope, on the leg to collect acceleration and rotation data when a person is walking. 
However, these approaches need to employ either a large number of sensors~\cite{anderson_shi_2019, tjhai2019using} or are evaluated using a high-cost IMU (over a few thousand dollars) \cite{kim_ju_park_2019}, which is not practical for daily use.
Therefore, there is a critical need for simple and intuitive models to estimate gait parameters using relatively few sensors. 

\begin{figure*}[t]
    \centering
    \includegraphics[width=0.7\linewidth]{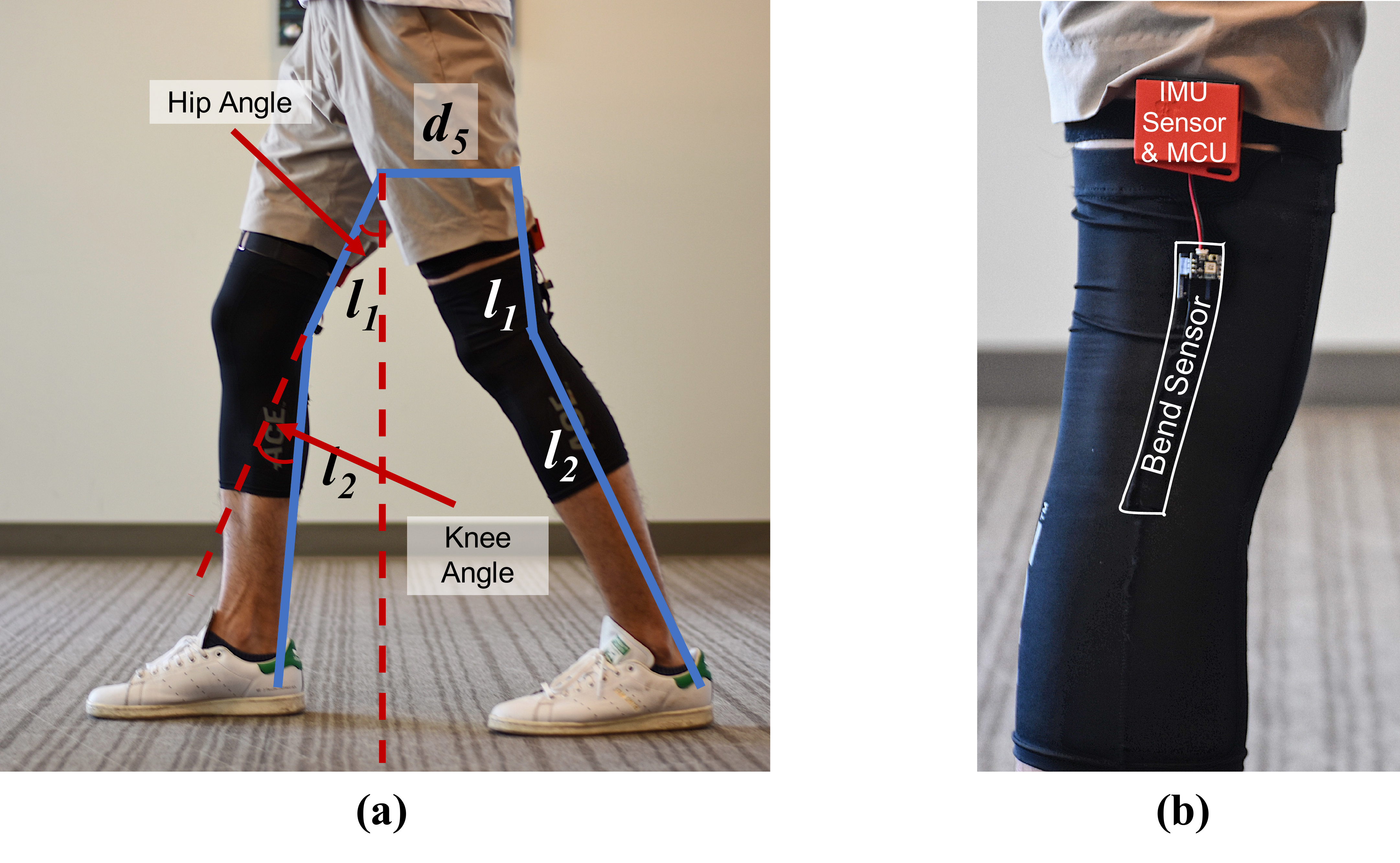}
    \caption{(a) Wearable setup used for gait analysis (b) Magnified view of the sensors}
    \label{fig:exp_setup}
    \vspace{-3mm}
\end{figure*}

This paper presents a wearable cyber-physical system (CPS), called \textit{MGait}, 
that combines physiological sensors, energy-efficient local processing, 
and real-time user feedback. 
We implement \textit{MGait} on a knee sleeve, as shown in Figure~\ref{fig:exp_setup} 
and demonstrate that it achieves the following goals:

\noindent \textbf{Functionality:} Enable continuous daily monitoring and \textit{real-time feedback} with 95\% average accuracy without relying on any clinical environment and experimental infrastructure.

\noindent \textbf{Power-performance:} 
Achieve real-time operation and mW-range operation using only one pair of IMUs and low-power bend sensors \textit{for the first time in literature}.

We meet these goals with two sensors per leg and commercial-off-the-shelf components that can be integrated with a total cost of less than \$160, 
in contrast to techniques that use sensors and equipment with over one order of magnitude higher cost.
A stretchable bend sensor mounted on the back of the knee measures the knee angle, while an IMU sensor above the knee measures the swing of the hip on each leg, as shown in Figure~\ref{fig:exp_setup}.
Data from the bend and IMU sensors are processed \textit{locally in real-time} to obtain these angles. 
We propose a novel biomechanical model and 
derive a \textit{closed-form expression} 
that computes the step length using the angle data. 
Using only our closed-form expression and sensor data leads to over 10\% estimation error due to sensor offset and measurement noise. We reduce this error significantly using a novel two-step nonlinear regression technique. 
Thorough experimental evaluations with seven subjects show that the proposed approach achieves a 5\% mean absolute percentage error (MAPE) with this optimization. 
In addition, we also present a recursive least square estimation technique that can tune the proposed model in the field. 
Finally, we employ the proposed model to derive other commonly used clinical metrics, including 
stance time, swing time, gait velocity, and stride length. 
The proposed system can be used to provide real-time feedback if any outlier behavior is observed, as illustrated in Section~\ref{sec:feedback}. 

We will release our empirical dataset to the public to enable further research in the gait analysis domain. In addition to the empirical dataset, we also implement a novel method to generate synthetic data. Specifically, we train a Conditional Generative Adversarial Nets (CGAN) using the empirical data to generate additional synthetic data. The synthetic data generation methodology will allow researchers to minimize the time and labor-consuming step of data collection. The data generated using the CGAN closely follows the patterns of the empirical with a mean deviation of 1$^\circ$.

\vspace{1mm}
In summary, this paper makes the following major contributions:

\begin{itemize} [leftmargin=*]
    \item An energy-efficient wearable cyber-physical system for real-time gait analysis and step length estimation.
    
    \item A novel analytical modeling approach and a closed-form expression for estimating step length, stride length, stance time, swing time, and gait velocity.
    
    \item A novel method to generate synthetic gait data using conditional generative adversarial nets.
    \item A thorough experimental evaluation, including offline and online estimation, with seven subjects and a new dataset~\cite{Github}.
\end{itemize}

In the following sections, Section~\ref{sec:related_work} reviews the related work. Section~\ref{sec:overview} introduces the gait analysis and the proposed approach. Section~\ref{sec:angle_est} and~\ref{sec:model_description} present angle and step length estimation approaches, respectively. Finally, Section~\ref{sec:experiments} presents the experimental results, Section~\ref{sec:synthetic_dataset_generation} proposes a technique to generate synthetic data based on CGAN, and  Section~\ref{sec:conclusion} concludes the paper.

\section{Related Work}\label{sec:related_work}
\raggedbottom

Gait analysis is vital for enabling a variety of CPS applications, such as feedback for patients under treatment, prediction of possible movement disorders~\cite{phillips2016fallpred, ferster2015fog}, athlete assessment, fall detection~\cite{piau2019fallpred}, and numerous other augmented and virtual reality applications. 
For this reason recent work has focused on using wearable sensors to estimate step length, which is a very important parameter in gait analysis~\cite{anderson_shi_2019,wang_sun_li_liu_2018,wu_wang_pottie_2015,kim_ju_park_2019,pepa_2016, tjhai2019using}.

Table~\ref{tab:rel_work} summarizes state-of-the-art step length estimation techniques that are closest to our work. 
Wang \textit{et al.} propose a biomechanical model that takes knee bending into account and uses data obtained from a total of four low power IMUs placed on both legs to estimate step length and gait asymmetry~\cite{wang_sun_li_liu_2018}. 
The proposed double pendulum model estimates the step length with a 5.5 cm root mean square error (RMSE).
This model considers the hip as a single hinge joint but ignoring the displacement of the hip joints in the direction of the movement thus, results in errors in step length estimation. 
Wu \textit{et al.} place two IMU dev-kits on the ankles of the subjects to estimate step length, using an inverted pendulum model with 3.69\% MAPE \cite{wu_wang_pottie_2015}. 
Their model does not consider knee and hip joints. Instead, it employs ankle-mounted motion sensors to calculate the step length using the leg length and the sine of the leg’s orientation. However, the error reported in their study is for the total walking distance, not for individual step lengths. 
Since positive and negative errors in individual step lengths cancel out, this reporting choice shows lower MAPE. For example, the proposed MGait technique MAPE decreased by 1\% to 2\% if we consider the total walking distance. Since the model assumes a single segment for each leg, it is able to calculate the step length using only two IMUs. 
Consequently, the power consumption of this approach is lower, compared to other more complicated approaches.
Their approach also requires a Kinect V2 setup to estimate leg lengths of the subjects in their study. We do not include it in the sensor type since it does not contribute to the model calculation.
Pepa \textit{et al.} utilize accelerometers in smartphones and develop an app to collect motion data. 
They also use an inverted pendulum model and estimate step length with less than 10\% MAPE \cite{pepa_2016}. 
None of these studies provide stride length and gait velocity estimates, which are two other main parameters used for gait function analysis \cite{pirker2017gait}. 

\begin{table*}[t]
\centering

\caption{A compilation of previous related work on gait analysis (``--'' means the results are not reported)}
\label{tab:rel_work}
\setlength{\tabcolsep}{4.5pt}
\scriptsize{
\begin{tabular}{@{}clllllllcccl@{}}
\toprule
Ref & \multicolumn{2}{l}{Step Length Err.} & \multicolumn{2}{l}{Stride Length Err.} & \multicolumn{2}{l}{Velocity Err.} & Sensor Type & \begin{tabular}[c]{@{}l@{}}Sensor \\ Count\end{tabular} & \begin{tabular}[c]{@{}l@{}}User \\ Feed.\end{tabular} & \begin{tabular}[c]{@{}l@{}}Wearable \\ Form-factor\end{tabular} & \begin{tabular}[c]{@{}l@{}}Power \\ (mW)\end{tabular} \\ \midrule
 & RMSE & MAPE & RMSE & MAPE & RMSE & MAPE &  &  &  &  &  \\
     \cmidrule(lr){2-3}
     \cmidrule(lr){4-5}
     \cmidrule(lr){6-7}
\cite{wang_sun_li_liu_2018} & 5.5 cm& - & - & - & - & - & MPU-6050 IMU & 4 & \ding{55} & \ding{51} & 36.10 \\
\cite{wu_wang_pottie_2015} & - & 3.69\% & - & - & - & - & \begin{tabular}[c]{@{}l@{}}MPU-9150 IMU\end{tabular} & 2 & \ding{55} & \ding{55} & 18.53\\
\cite{pepa_2016} & - & <10\% & - & - & - & - & Smartphone accel. & 1 & \ding{55} & \ding{55} & - \\
\cite{anderson_shi_2019} & 4.1 cm& 4.75\% & 6.3 cm & 3.70\% & 0.07 m/s & 4.23\% & \begin{tabular}[c]{@{}l@{}}\textit{High Cost} UWB, \\ MPU-6050 IMU\end{tabular} & 2+2 & \ding{55} & \ding{51} & 145.20 \\
\cite{tjhai2019using} & - & <5\% & - & <5\% & - & - & \begin{tabular}[c]{@{}l@{}}MPU-6050 IMU,\\ MPU-9250 IMU\end{tabular} & 3+4 & \ding{55} & \ding{51} & 64.13\\
GR* & - & <2\% & - & <1\% & - & <1\% & GR* Proprietary & >10k & \ding{55} & \ding{55} & - \\
\rowcolor{lightgray}
\begin{tabular}[c]{@{}l@{}}\textbf{\textit{MGait}} \\\end{tabular} & \textbf{4.3 cm} & \textbf{5.47\%} & \textbf{6.24 cm} & \textbf{3.90\%} & \textbf{0.03 m/s} & \textbf{2.33\%} & \textbf{\begin{tabular}[c]{@{}l@{}}MPU-9250 IMU, \\ Bendlabs 1-Axis\end{tabular}} & \textbf{2+2} & \textbf{\ding{51}} & \textbf{\ding{51}} & \textbf{16.97} \\ \bottomrule
\end{tabular}
}\begin{flushleft}
{\footnotesize GR*: GAITRite \cite{gaitrite}}
\end{flushleft}

\vspace{-5mm}
\end{table*}

A recent work addresses the limitations in the previous studies, by using four \textit{high-cost ultra-wideband} (UWB) \textit{distance sensors} and four IMUs on the back and front of both feet \cite{anderson_shi_2019}. 
It employs a 
geometrical trapezoid distance model and estimates step length with 4.1 cm RMSE and 4.75\% MAPE. 
However, the high processing requirements of the data from eight sensors increase the complexity and the power consumption of the system. 
Another recent study places two IMUs on the feet, two IMUs on the shanks, two IMUs on the thighs, and one IMU on the pelvis, using a total of seven IMUs~\cite{tjhai2019using}.
It uses a Kalman filter framework to calculate the orientation of each sensor.
Since they use seven IMUs, the proposed approach is not practical, and the power consumption of the system is high, due to a large number of sensors and their processing cost.

In summary, none of these studies targets real-time feedback to the user 
and considers the system power consumption.
In contrast to all of these approaches, \textit{MGait} estimates the key gait parameters 
with the minimum number of IMUs and low-power bend sensors (one on each leg), 
and a low power micro-controller using a novel closed-form expression. 
It achieves accurate step length, stride length, and gait velocity.
Low power consumption and processing requirements of \textit{MGait} can enable 
the first low-cost, low power, and wearable CPS for gait analysis with user-feedback functionality.

\section{Overview of Gait Cycle and \textit{MGAIT} Approach} \label{sec:overview}

\subsection{Gait Cycle Definition and Segmentation}

Accurate step length modeling requires a clear understanding of the periodic gait cycle and two key angles~(\textit{All the angles and the length of the limbs in the following text are considered on the sagittal plane unless otherwise specified.}):

\vspace{-0.5mm}
\begin{definition}
\textit{Knee Angle}~($\beta$) is the angle formed at the joint of the thighs and legs, as shown in Figure~\ref{fig:exp_setup}(a). 
It ranges between 0$\degree$ (when the leg is straight) 
and $\approx$50$\degree$ (during the swing).
\end{definition}

\begin{figure}[t]
    \centering
    \includegraphics[width=1\linewidth]{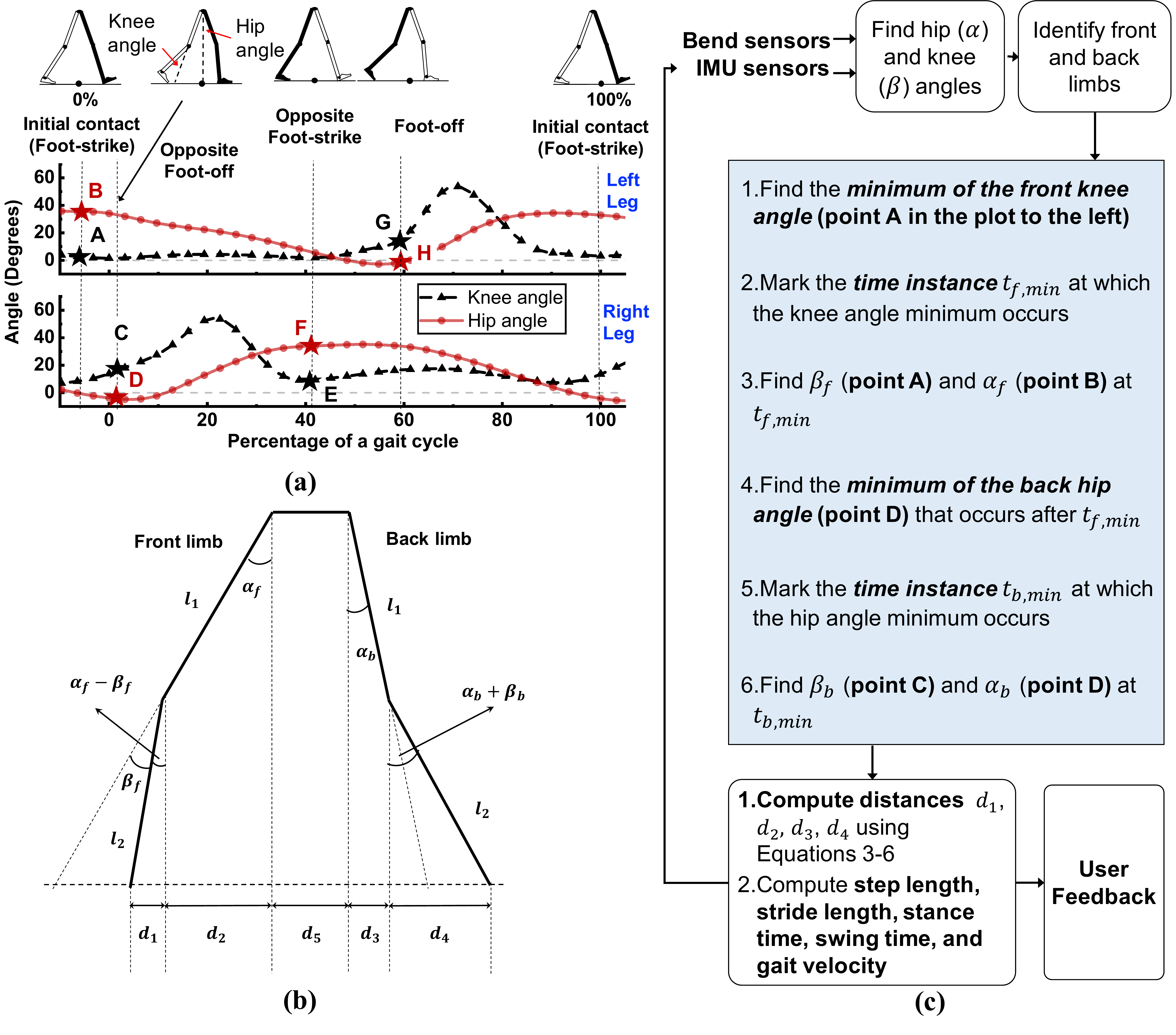}
   \caption{ (a) Overview of the gait cycle. This figure is generated using the wearable setup shown in Figure~\ref{fig:exp_setup} (b) Illustration of the step length components used in our model. These components, defined in Equations~\ref{Eq:d1}--\ref{Eq:d3} are added to get the total step length. (c) Overview of the proposed approach. Using the hip and knee angles, we first identify the points of interest shown in part (a).Using the angles at the points of interest, we calculate the step length components using Equations~\ref{Eq:d1}--\ref{Eq:d3} to obtain the total step length.}
    \label{fig:gait}
\end{figure}

\vspace{-0.5mm}
\begin{definition}
\textit{Hip Angle}~($\alpha$) is the angle formed by the inner thigh and the vertical axis, as shown in Figure~\ref{fig:exp_setup}(a). 
It is positive ($\approx$40$\degree$) when the thigh is in front of the torso and negative when the thigh is behind the torso, as shown in Figure~\ref{fig:gait}(a).
\end{definition}

Knee and hip angles change periodically during the gait cycle. For the front limb, the hip and knee angles are denoted by $\alpha_f$ and $\beta_f$, respectively. Similarly, the hip and knee angles of the back limb are denoted using $\alpha_b$ and $\beta_b$, respectively. The details of how to find front and back limb using key events are explained in Section~\ref{sec:algorithm}.
There are four key events and two stances~\cite{carollo2002strategies, shi2019effect}, 
as marked in Figure~\ref{fig:gait}(a):
\begin{enumerate} [leftmargin=*]
    \item \textit{Initial Contact}: At the beginning of the gait cycle, the foot just touches the ground. The knee angle of the front leg (the left leg shaded in Figure~\ref{fig:gait}(a)) reaches its first minimum and the hip angle is close to its maximum, as shown with markers A and B, respectively.

    \item \textit{Opposite foot-off}: The opposite leg (the right leg in Figure~\ref{fig:gait}(a)) starts lifting from the ground. After this point, the right leg is in a forward swing, until it strikes the ground. The angles in this stance are denoted by markers C and D.
    
    \item \textit{Opposite foot-strike}: At this point, the right leg completes the swing (i.e., one complete step) and touches the ground. 
    The right leg is in front of the torso while the left leg is behind the torso. Thus, the knee angle of the right leg is at a minimum, as shown with the marker E in Figure~\ref{fig:gait}(a). 

    \item \textit{Foot-off}: The hip angle of the left leg reaches its minimum value. The left also starts its swing forward for the next step. This stance is shown using markers G and H.
    
\end{enumerate} 

These four events of one leg belong to its stance phase. The period from foot-off to the next initial contact is the swing phase.

\subsection{Flow of the MGait Approach}
The goal of the \textit{MGait} approach is to model and estimate the step length and relevant parameters using the setup shown in Figure~\ref{fig:exp_setup}(a).
We place a bend sensor vertically aligned to the backside of the knee when the subject is standing. To measure the knee angle with the best accuracy, the midpoint of the bend sensor is adjusted at the knee level. 
The IMU is placed on the hip, perpendicular to the ground when the subject is standing.
Accelerometer and gyroscope data from the IMU are used to calculate the hip angle.
We process the angle data to find the major events in the gait cycle, as outlined in Figure~\ref{fig:gait}(c).
Specifically, we find the points \{A, B, C, D\} for the first step and \{E, F, G, H\} for the second step plotted in Figure~\ref{fig:gait}(a). 
The angles at these points are used to calculate the step length components $d_1$--$d_4$ shown in Figure~\ref{fig:gait}(b). These components are used by the \textit{MGait} approach to compute the step length, stride length, and gait velocity. The details of the analytical model to compute the step length, stride length, and gait velocity are presented in Section~\ref{sec:model_description}.

\section{Hip and Knee Angle Estimation}\label{sec:angle_est}


\subsection{Sensor Calibration and Data Pre-processing}
\vspace{-0.5mm}
Due to changing environment, sensor data is noisy and drift prone. For instance, angle measured by the bend sensor for the resting state can vary with time and IMU sensors may have offsets.
To calibrate the sensors, we instruct the users to stand still with a straight leg for a few seconds before the experiment. 
During this period, we record the sensor readings and find their median 
values as the offsets. We then subtract the offsets from the sensor data during actual use. For example, the self-calibration ensures that the IMU measures only the gravity when the user stands in a straight position.

After calibrating the sensors, we apply a downsampling and smoothing filter to reduce noise in the data. For each sensor, we obtain the downsampled and smoothed samples as:
\begin{equation} \label{Eq:subsample}
s_s[Mk] = \frac{1}{2M}\sum_{i = M(k-1)+1}^{M(k+1)} s[i]
\end{equation}
where $s_s$ is the smoothened data stream, $M$ is the downsampling factor, $k$ is the sample index, and $s$ is the data stream before filtering. 
The IMU and bend sensor sampling rates are 250~Hz and 100~Hz, respectively. 
Since the fastest gait is less than 10 Hz, we downsample the
data to 25~Hz, which is sufficient, considering the Nyquist rate.
The specifications of sensors before and after filtering are shown in Table~\ref{tab:sensorspecifications}.

\begin{table}[h]
\centering
\caption{Sensors specifications}
\label{tab:sensorspecifications}
\begin{tabular}{@{}llll@{}}
\toprule
               & \begin{tabular}[c]{@{}l@{}}Sensor\\ range\end{tabular} & \begin{tabular}[c]{@{}l@{}}Sampling \\ rate (Hz)\end{tabular} & \begin{tabular}[c]{@{}l@{}}Sampling rate after \\ median filter (Hz)\end{tabular} \\ \cmidrule(l){2-4} 
Accelerometer  & ±8g                                                    & 250                                                           & 25                                                                                \\ \midrule
Gyrometer      & ±250dps                                                & 250                                                           & 25                                                                                \\ \midrule
Bending sensor & ±180°                                                  & 100                                                           & 25                                                                                \\ \bottomrule
\end{tabular}
\end{table}

%

\vspace{-1mm}
\subsection{Knee Angle Computation and Visualization}
\begin{figure}[h]
    \centering
    \includegraphics[width=0.60\linewidth]{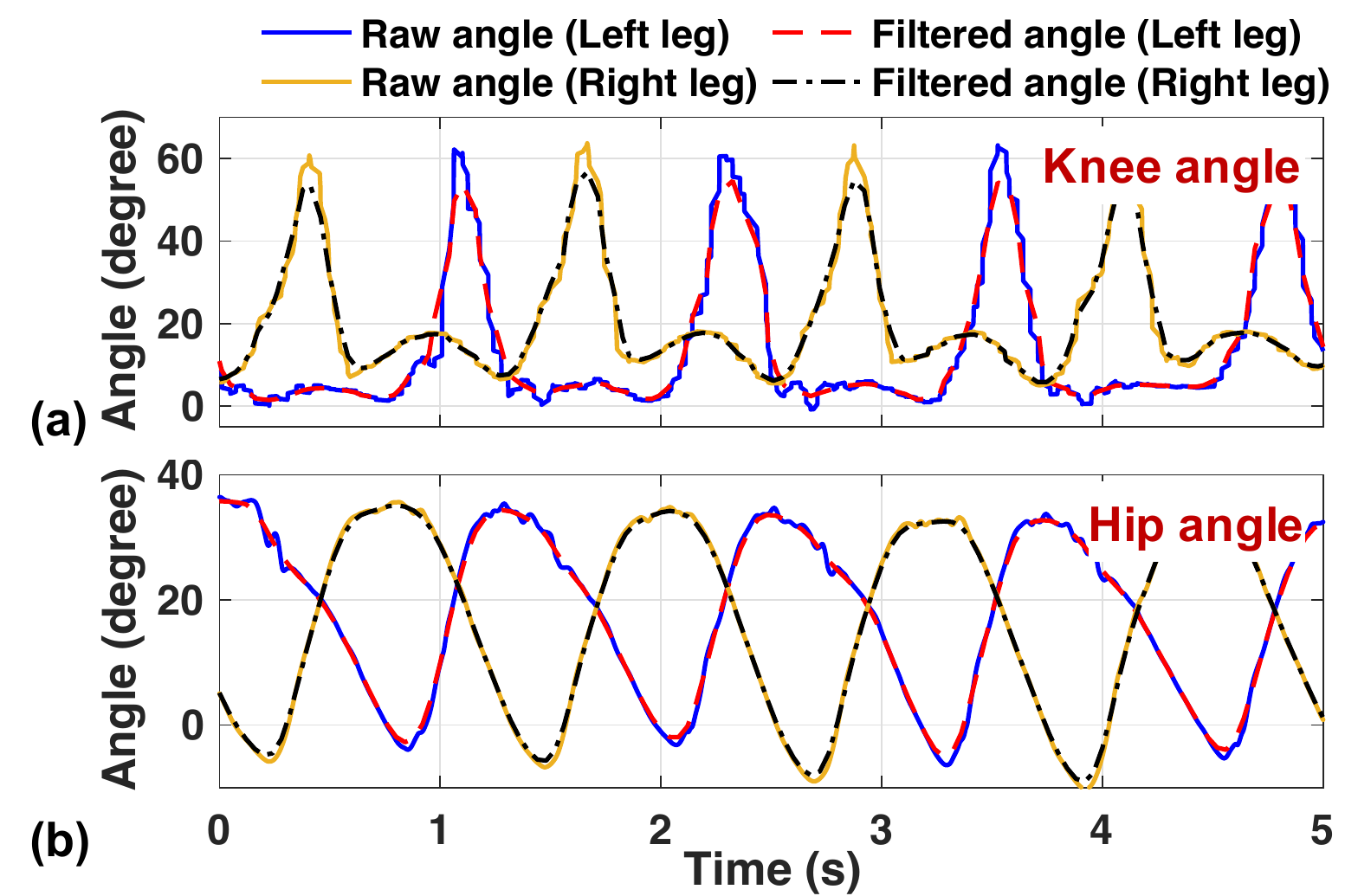}
    \caption{Visualization of raw and median filtered (a) knee angle and (b) hip angle of both legs during walking}
    \label{fig:knee}
\end{figure}

\vspace{-0.5mm}
The output of the bend sensor (located at the back of the knee) 
is the angle displacement experienced by the sensor, as illustrated in Figure~\ref{fig:exp_setup}(a). 
Sample knee angle data from our experiments is visualized in Figure~\ref{fig:knee}(a). 
A comparison between the raw and filtered data shows that the averaging filter smooths out the variations in the data. 
The knee angles reach a maximum of about 50--55\textdegree~as the leg swings. 
Another key observation is that the angles of the two legs exhibit approximately 50\% phase shift when walking. This is expected for healthy subjects, since their steps are of similar length. 

\vspace{-1mm}
\subsection{Hip Angle Computation and Visualization}

We find the hip angle using the IMU sensor located right above the knee, as shown in Figure~\ref{fig:exp_setup}(a). 
There are several approaches to filter the accelerometer and gyroscope samples from the IMU and obtain the hip angle. Among these, we compared the performance of the complementary filter~\cite{higgins1975comparison} and the Madgwick filter~\cite{madgwick2011estimation}, which are the most commonly used techniques. 
Since the Madgwick filter outperforms the complementary filter in terms of the ability to converge and the smoothness, 
we used it in the final implementation. 
The Madgwick filter uses a quaternion representation of the accelerometer samples and applies a gradient descent algorithm to calculate the error in the direction of the gyroscope samples. By compensating for the error, it accurately estimates the orientation of the IMU even during motion, without being affected by gyroscopic drift~\cite{madgwick2011estimation}.

\begin{figure}[t]
    \centering
    \includegraphics[width=0.55\linewidth]{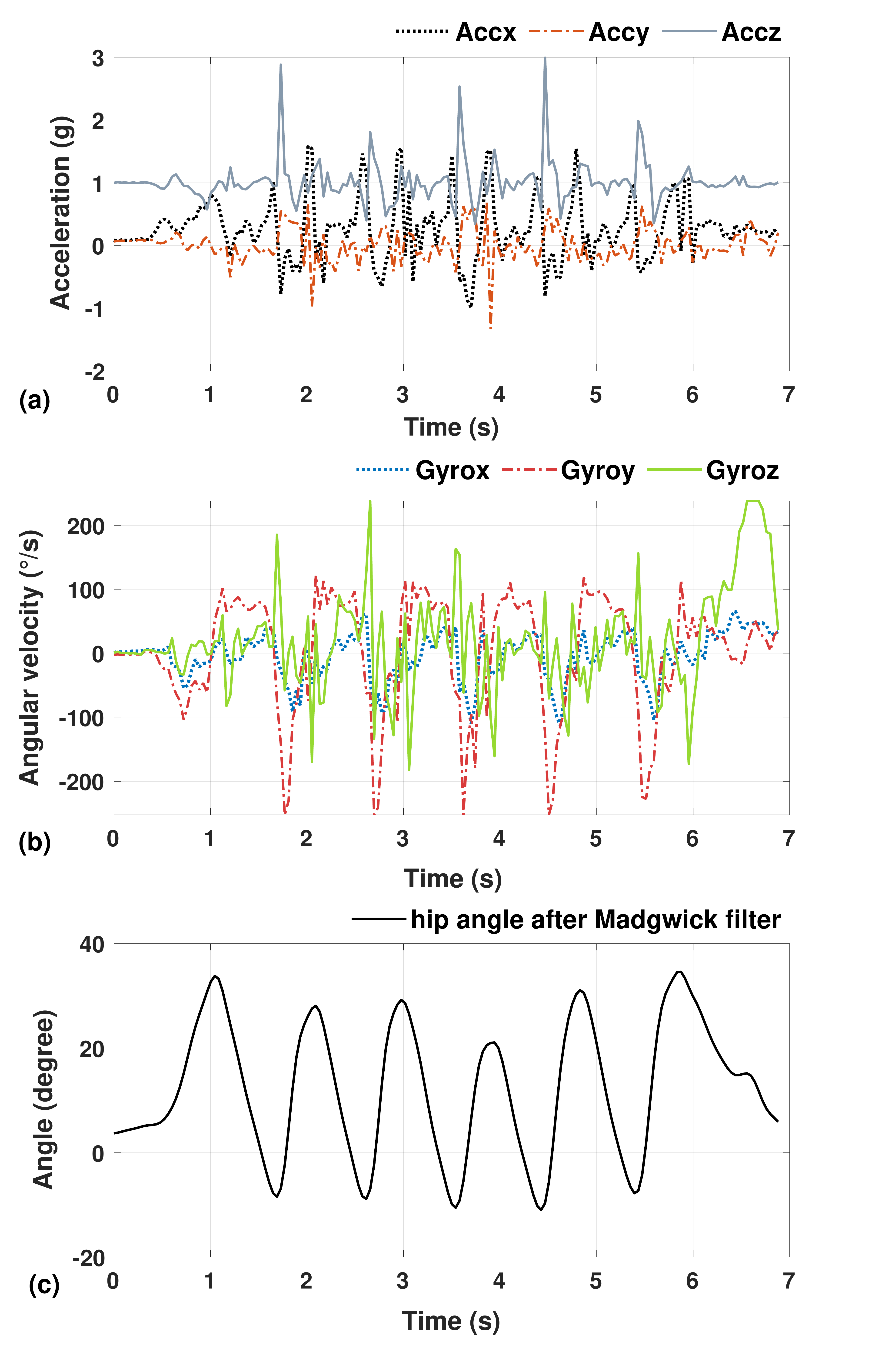}
    \vspace{-4mm}
    \caption{Visualization of acceleration values (a) and angular velocity values (b) forming hip angle (c) after the Madgwick filter}
    \vspace{-4mm}
    \label{fig:rawdatatohip}
\end{figure}

Figure~\ref{fig:knee}(b) plots left and right hip angles during walking. 
The hip angles vary between a minimum of about -10\textdegree~to a maximum of 35\textdegree~and exhibit a 50\% phase difference, as expected. We also show inputs and outputs of the Madgwick filter in Figure~\ref{fig:rawdatatohip}. Specifically, the Madgwick filter takes the accelerometer and gyroscope data shown in Figures~\ref{fig:rawdatatohip}(a) and \ref{fig:rawdatatohip}(b) as inputs. Using these inputs, it generates the smoothed hip angles, as shown in \ref{fig:rawdatatohip}(c). 

\vspace{-0.5mm}
\subsection{Finding the Key Events during the Gait Cycle}
\vspace{-0.5mm}
\label{sec:algorithm}

We need to identify the key gait cycle events, such as initial contact and foot-off, 
to convert the knee and hip angles into step length. 
From our gait cycle analysis, we know that these events occur when the knee angle is at a minimum for one leg (point \textbf{A} in Figure~\ref{fig:gait}(a)), 
while the hip angle is at a minimum for the other (point \textbf{D} in  Figure~\ref{fig:gait}(a)).
To obtain the minimum values, we continuously monitor the five-point derivative of the two angles. 
A minimum is marked whenever the derivative changes from negative to positive or zero to positive. 
After detecting the minimum of the knee angle, 
we designate the corresponding leg as the front limb for the current step. 
We then call the knee angle $\beta_f$ and hip angle $\alpha_f$ for the front limb.
For example, the knee angle $\beta_f$ and hip angle $\alpha_f$ in Figure~\ref{fig:gait}(a) correspond to points \textbf{A} and \textbf{B}.  
We then wait for the minumum of the hip angle to occur in the other leg, which is referred to as the ``back limb''. 
At this point, the knee angle $\beta_b$ and the hip angle $\alpha_b$ for the back limb corresponds to points \textbf{C} and \textbf{D} in Figure~\ref{fig:gait}(a). 
The angle values are then plugged into our analytical model to estimate the step length, as detailed in Section~\ref{sec:model_description}. 
This process is repeated continuously, where front and back limbs alternate.
\section{Proposed Gait Analysis Model} \label{sec:model_description}
\vspace{-0.5mm}

\subsection{Parameter Definitions}
\textit{MGait} estimates the step length, stride length, and gait velocity defined below in real-time. 

\begin{definition}
\textit{Step Length} is the distance between the front and back feet, when the front limb is in the initial contact stance and the back limb is in the foot-off stance. 
\end{definition}
\vspace{-0.5mm}
\begin{definition}
\textit{Stride Length} is the distance between two foot strike stances of the same leg. 
\end{definition}
\begin{definition}
\textit{Stance Time} stands for the period that from the foot touches the ground to the same foot leaves the ground.
\end{definition}
\vspace{-0.5mm}
\begin{definition}
\textit{Swing Time} stands for the period that from the foot leaves the ground to the same foot touches the ground.
\end{definition}

\begin{definition}
\textit{Gait Velocity} is the ratio of the stride length and the time taken to complete one stride.

\end{definition}

\subsection{Gait Parameter Estimation}
\begin{figure}[b]
    \centering
    \includegraphics[width=0.65\linewidth]{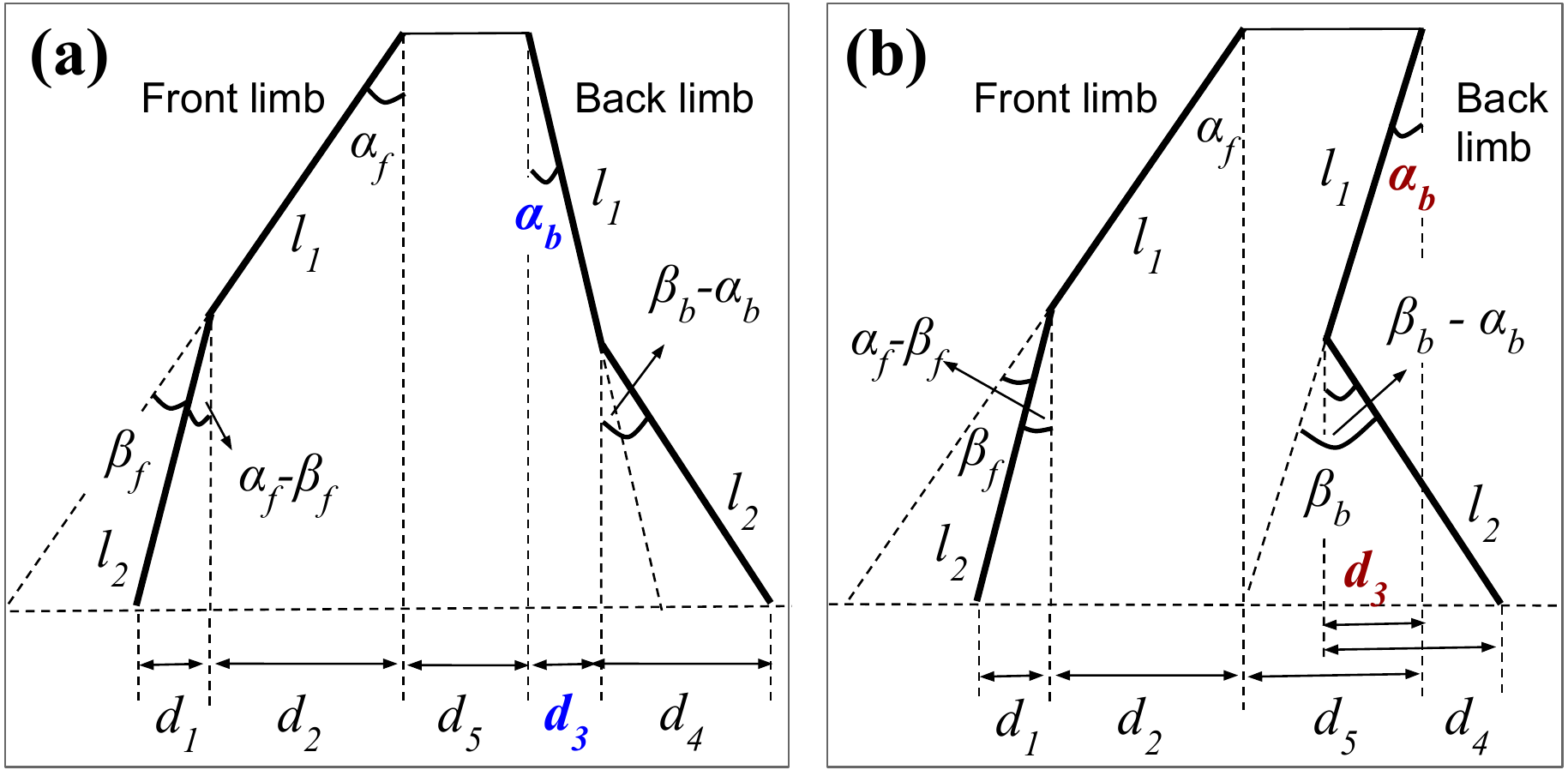}
    \caption{Stick diagram for step length calculation. The stick model has two cases, depending on the position of the back limb. (a) upper part of the back limb is behind the torso and (b) upper part of the back limb is parallel or slightly in front of the torso. The figures show the magnitude of all the angles in the model.}
    \label{fig:model1}
\end{figure}

We represent the stances during a step by the geometric stick diagram shown in Figure~\ref{fig:model1}. 
The left side of each drawing shows the front limb in the initial contact stance, while the right side shows the back limb in the foot off position. 
There are two cases of the stick diagram, depending on the position of the back limb. 
In the first case shown in Figure~\ref{fig:model1}(a), the back limb is extended behind the torso of the subject. 
In contrast, the upper part of the back limb aligns with the torso of the subject
in the second case shown in Figure~\ref{fig:model1}(b). 
We consider these two cases separately, since they change how the various components of the step length contribute to the model.

The step length is a function of the length of the leg of the subject. 
The length of the leg between the gluteus (hip) and popliteus point (knee joint) is denoted by $l_1$, as shown in Figure~\ref{fig:model1}(a). 
The length of the leg from the popliteus to the calcaneus point (ankle) is given by $l_2$. We then consider the knee and hip angles formed by both legs in the stick model. 
Using these definitions, we can express the step length as a sum of five components shown in Figure~\ref{fig:model1}.
Next, we describe the definitions of each of these components.

\noindent\textbf{$d_1$:} The $d_1$ component is the projection of the front leg on the ground. Figure~\ref{fig:model1}(a) shows that the angle formed by the vertical line and the leg is $\alpha_f - \beta_f$. Using this angle, the value of $d_1$ is written as:
\begin{equation} \label{Eq:d1}
    d_1 = l_2\sin(\alpha_f-\beta_f)
\end{equation}

\noindent\textbf{$d_2$:} $d_2$ is the projection of the front thigh to the ground. Since the angle formed by the thigh and the vertical axis is given by $\alpha_f$, the projection is:
\begin{equation} \label{Eq:d2}
    d_2 = l_1\sin(\alpha_f)
\end{equation}

\noindent\textbf{$d_3$:} $d_3$ is the projection of the back thigh to the ground. Since the angle formed by the thigh and the vertical axis is given by $\alpha_b$, the projection is:
\begin{equation} \label{Eq:d3}
    d_3 = l_1\sin(-\alpha_b)
\end{equation}
There is a negative sign in Equation~\ref{Eq:d3} to account for the two cases shown in Figure~\ref{fig:model1}(a) and Figure~\ref{fig:model1}(b). In Figure~\ref{fig:model1}(a), the $d_3$ has to be added to obtain the total step length. Therefore, we must ensure that the sign of $d_3$ is positive. By our definition of the hip angle, $\alpha_b$ is negative in Figure~\ref{fig:model1}(a) because it is behind the torso. The inclusion of a negative sign in Equation~\ref{Eq:d3} ensures that $d_3$ is positive. 
Conversely, in Figure~\ref{fig:model1}(b) $d_3$ overlaps with $d_4$ and $d_5$. Consequently, it has to be subtracted from the step length. In this case, $\alpha_b$ is positive because it is in front of the torso. Hence, we obtain a negative sign for $d_3$, ensuring its subtraction from the total step length.

\noindent\textbf{$d_4$:} $d_4$ is the projection of the back leg to the ground. Similar to $d_1$, we first need to calculate the angle between the back leg and the vertical line from the knee to the ground. In case of Figure~\ref{fig:model1}(a), the magnitude of this angle is given by the sum of angles $\alpha_b$ and $\beta_b$, whereas for Figure~\ref{fig:model1}(b) the magnitude is given by the difference of $\beta_b$ and $\alpha_b$. The sign of $\beta_b$ is always positive, while the sign of $\alpha_b$ is negative in Figure~\ref{fig:model1}(a) and positive in Figure~\ref{fig:model1}(b). Therefore, when we consider the signs of the respective angles, the resulting angle between the back leg and the vertical line becomes $(\beta_b - \alpha_b)$. Hence, the projection is expressed as:
\begin{equation}
    d_4 = l_2\sin(\beta_b-\alpha_b)
\end{equation}
\noindent\textbf{$d_5$:} $d_5$ is given by the diameter of the subject's thigh below the gluteus. It is included, since the stick diagram does not cover the width of the user's leg. It is measured at the beginning to personalize the model to each user.

\noindent Adding $d_1$--$d_5$ the total step length is obtained as:
\begin{equation}\label{eq:distance}
    D = l_2\sin(\alpha_f-\beta_f) + l_1\sin(\alpha_f) + l_1\sin(-\alpha_b) + l_2\sin(\beta_b-\alpha_b) + d_5
\end{equation}
After obtaining the step length, we calculate the stride length by adding the lengths of two consecutive steps. 
We compute the gait velocity by dividing the length of a sequence of five strides by the time. The stride time is obtained by subtracting two consecutive timestamps obtained when finding the minimums, as described in Figure~\ref{fig:gait}(b).

\subsection{Real-time User Feedback} \label{sec:feedback}
Gait analysis is commonly used in patient rehabilitation and health monitoring. Therefore, providing feedback to the users about any abnormalities in gait is a crucial aspect of gait analysis. To this end, we provide the following feedback to the user in real-time, as shown in Figure~\ref{fig:gait}(b). Note that real-time feedback is optional, and users can always choose offline analysis of the gait after their trial. Some users with functional disabilities might find it uncomfortable if the buzzer or the LED is always on. Therefore, we enable logging the gait data during the whole trial and provide the batch least square algorithm  users can use to analyze their gait patterns offline.
We emphasize that the proposed feedback mechanism is tested only with healthy subjects mimicking the limping behavior due to difficulty in accessing actual patients. Nevertheless, these studies can pave the way for applying the proposed algorithms to real patients. As our future work, we plan to collect the users’ feedback, improve the framework’s robustness, and test it with actual patients, e.g., those suffering from the freezing of gait (FoG).

\noindent\textbf{Gait Asymmetry Detection}: Patients with movement disorders often have different left and right step lengths \cite{yogev2006gaitasym}.
Therefore, we include gait asymmetry as one of the feedbacks that we provide to the user. We notify the user of gait asymmetry using a buzzer on our device. The buzzer sounds an audible alarm to the user such that they can take appropriate action to correct the gait asymmetry. 



\noindent\textbf{Gait Velocity Reduction Detection:}
Falling is among the primary causes of death in the elderly population \cite{piau2019fallpred}. 
Change in the gait velocity over a long period is considered as a significant cue to predict future falls \cite{piau2019fallpred, phillips2016fallpred}. \textit{MGait} keeps track of the gait velocity of each patient and notifies the patient or the healthcare provider, if there is a specific trend in the decrease of gait velocity that can be identified as a potential risk of fall.
    
\noindent\textbf{Step/Stride Length Reduction Detection:} 
Reduced stride/step length is one of the most prominent features of various movement/motor disorders in the elderly population, especially in Parkinson's Disease (PD) patients~\cite{brognara2020beneficial}.
Therefore, we keep track of the user's moving average step/stride length and notify the user if there is a significant reduction in step length using an LED on our device. 

\vspace{2mm}
\noindent\textbf{Feedback Algorithm:}
Gait asymmetry and step/stride length reduction can be detected by observing the trend of step length over time. In a healthy walking pattern, left-to-right step length and right-to-left step length are similar in length. In contrast, when gait asymmetry and step/stride length reduction are present, one step is shorter than the other. That is, the difference between left-to-right and right-to-left step length is larger than healthy walking. 
Using the insights, we design the feedback algorithm using the percentage gait asymmetry in strides.

The percentage gait asymmetry is employed as a metric to determine if a stride is asymmetric. The first step of our algorithm is to calculate the percentage gait asymmetry in a stride. The percentage gait asymmetry is defined in Equation~\ref{eq: gaitasym}:
\begin{equation}
\label{eq: gaitasym}
Gait_{asym}^{i} = \frac{\left | L_{left}^{i} - L_{right}^{i} \right |}{0.5 \times (L_{left}^{i} + L_{right}^{i} )} \times 100\%
\end{equation}
, where $i$ is the stride count and $L$ is the step length of a stride. $L_{right}^{i}$ represents the $i_{th}$ right step length, and the similar explanation applies for the left situation. The percentage gait asymmetry quantifies the difference between the left and right steps length in the same stride. If the difference exceeds a threshold, we then identify that this is an asymmetric stride. The value of the difference threshold has to be chosen carefully to avoid false positives. Since percentage gait asymmetry involves the normalization operation itself, it accurately reveals the percentage left-to-right step length difference even for subjects with different height and static parameters~(l1, l2, and d5). Our dataset shows that the percentage gait asymmetry in normal step length when walking is less than 20\%. To avoid extra false alarms, in our implementation, we choose the threshold as 25\%.
At the same time, all steps length and step speed data are logged. By using the proposed algorithm, the users can analyze their gait patterns offline.

\vspace{2mm}
\section{Experimental Evaluation} \label{sec:experiments}

\subsection{Experimental Setup}
\noindent \textbf{Wearable Device:}
\textit{MGait} framework uses the wearable setup shown in Figure~\ref{fig:exp_setup}(a). The magnified view of the wearable setup is shown in Figure~\ref{fig:exp_setup}(b).
It consists of a wearable bend sensor~\cite{bendlabs} and a Texas Instruments CC2650 Sensortag~\cite{ticc2650}. We sample the bend sensor and sensortag at 100 Hz and 250 Hz, respectively.
The proposed gait analysis technique is implemented on the TI Sensortag to enable runtime analysis and user feedback.


\noindent\textbf{User Studies:} We collected data from a total of seven subjects (S1--S7), 
following an official protocol approved by the IRB board of our institution. The information, including the static parameters ($l_1$, 
$l_2$, and $d_5$), height, age, and gender of 7 subjects, are shown in Table~\ref{tab:subjectinfo}. 
Each subject participated in six trials. Four of the six trials were regular free walking with normal pace, whereas in the remaining two the subjects were asked to imitate limping, 
resulting in a total of 806 steps. In addition to the empirical data, we also proposed a method to generate additional data, as described in Section~\ref{sec:synthetic_dataset_generation}. The synthetic data generation augments the data collected from the 7 subjects to create a richer dataset for further research. Both empirical and synthetic data will be released to the public, along with this paper.

Accurate step length reference is critical to evaluating \textit{MGait}. For most of the experiment, subjects walked on a 7-meter long white paper roll. We rubbed the bottom of the subject's shoes with washable ink.
The marks left on the paper are then used to capture the user steps. After each trial, the distances between the marks on the paper are recorded as step lengths. 
For the last two subjects, we employed the GAITRite~\cite{gaitrite} system to obtain the step length reference. 

Across all the experiments, the recorded ranges for step length, stride length, and gait velocity are 25--78 cm, 69--156~cm, and 0.60--1.27~m/s respectively. We also measure the stance time and the swing time of our dataset. Stance time of gait stands for the period that from the foot touches the ground to the same foot leaves the ground. Swing time of gait stands for the period that from the foot leaves the ground to the same foot touches the ground. The dataset's stance time and swing time ranges are 0.38–-0.75~s and 0.23–-0.54~s, respectively. Generally, 60\% of one gait period is stance time, while 40\% is swinging time. The step length distribution in our experimental data is as follows. 80\% of the step lengths are between 60 cm and 70 cm, 9\% of them fall into 50 cm to 60 cm range, 6\% are shorter than 50 cm, finally and 5\% of them are longer than 70 cm. Similarly, 76\% of the stride lengths are from between 120 cm and 140 cm, 11\% of them vary from 100 cm to 120 cm, 7 of them are shorter than 100 cm, and 6\% of them are longer than 140 cm. We also perform a statistical analysis on our dataset. For each subject's step length and stride length, we calculate the mean and standard deviation. Moreover, a 95\% confidence interval is computed. The detailed results are shown in Table~\ref{tab:subjectinfo}.


\begin{table}[h]
\centering
\caption{Overview of the dataset for each subject. std represents standard deviation and CI represents confidence interval.}
\label{tab:subjectinfo}
\setlength{\tabcolsep}{2.5pt} 
\begin{tabular}{@{}lccccrcrc@{}}
\toprule
  &\multirow{2}{*}{\begin{tabular}[c]{@{}c@{}}Static param. \\ l1/l2/d5 (cm) \end{tabular}} & \multirow{2}{*}{\begin{tabular}[c]{@{}c@{}} Height \\ (cm) \end{tabular}} & \multirow{2}{*}{Age} & \multirow{2}{*}{Gender} & \multicolumn{2}{c}{Step length (cm)} & \multicolumn{2}{c}{Stride length (cm)} \\
 &                    &                    & & &  Mean $\pm$ std           & CI          & Mean $\pm$ std   & CI  \\ \midrule
S1 & 30/45/14               & 193        & 23  & M      &61.33 ± 6.65 & 60.31 - 62.35 & 122.87 ± 11.07 & 120.46 - 125.29\\
S2 & 25/40/15               & 175        & 26  & M      &61.75 ± 13.39 & 57.67 - 65.82 & 123.50 ± 25.41 & 112.23 - 134.76\\
S3 & 29/40/16               & 188        & 26  & M      &73.37 ± 3.83 & 72.10 - 74.65 & 146.66 ± 6.20 & 143.58 - 149.74\\
S4 & 27/33/12               & 170        & 29  & M      &58.21 ± 13.56 & 53.98 - 62.44 & 116.42 ± 25.33 & 104.89 - 127.96\\
S5 & 20/25/15               & 163        & 26  & F      &48.05 ± 6.58 & 46.42 - 49.68 & 96.65 ± 11.57 & 92.48 - 100.82\\
S6 & 28/40/16               & 181        & 25  & M      &60.47 ± 2.89 & 60.19 - 60.75 & 121.01 ± 5.29 & 120.28 - 121.74\\
S7 & 22/28/22               & 158        & 25  & F      &57.76 ± 5.35 & 56.14 - 59.39 & 115.53 ± 10.36 & 110.4 - 120.13\\ \bottomrule
\end{tabular}
\end{table}




\subsection{Offline and Online Estimation of Static Parameters}

The proposed model has three user-specific static parameters:
length of the thigh ($l_1$), length of the leg ($l_2$), and hip diameter ($d_5$). 
Since measuring them is subject to human error, we employ both offline batch processing and online least squares regression models to determine these parameters. 

\label{sec:est_stat_param}
\subsubsection{Batch Least Square Estimation}
\label{sec:est_stat_param_batch}

Batch processing is a common method used in the least square estimation. 
For our step length estimation problem, batch processing is an effective solution when the gait analysis is performed offline. For instance, a patient can walk for a few minutes, while measuring the golden step length reference. 
The collected data then can be used to reduce the measurement error by tuning the model.

We employ a nonlinear least square estimation to solve the problem. First, we measure their nominal values by hand and 
record them as $l_1^{nom}, l_2^{nom}, d_5^{nom}$. We then use our test data and these nominal values to formulate the following regression problem:
\vspace{-1mm}
\begin{align}\nonumber
& \text{minimize} & \sum_{i=1}^{N} ||D_{ref,i} - D_i(l_1,l_2,d_5) ||^2    \\\nonumber
& \text{subject to} & 0.9\times l_1^{nom} \leq l_1 \leq 1.1 \times l_1^{nom} \\ \label{eq:opt1}
& & 0.9 \times l_2^{nom} \leq l_2 \leq 1.1 \times l_2^{nom} 
\\ \nonumber
& & 0.9 \times d_5^{nom} \leq d_5 \leq 1.1 \times d_5^{nom}
\end{align}
where $N$ is the number of steps used for regression and $D_{ref,i}$ is the step length reference for $i^{\mathrm{th}}$ step. 
The objective in Equation~\ref{eq:opt1} minimizes the sum of squared error between $D_{ref,i}$ and our estimate $D_i$ obtained with  Equation~\ref{eq:distance} using the measured angles ($\alpha_f$, $\alpha_b$, $\beta_f$, $\beta_b$). 
We constrain the optimization variables $l_1$, $l_2$, and $d_5$ within 
10\% of their nominal values to ensure that they do not overfit to unrealistic values.
As an example, the parameters of S1 are 30.0, 45.0, and 12.0 cm initially. After the regression, 
they are corrected as 33.0, 42.1, and 13.2 cm, respectively. 
\textit{This method is used once for each subject to find their static parameters.}

\subsubsection{Recursive Least Square Estimation}
\label{sec:est_stat_param_rls}
It is also useful to fine-tune the model in real-time after it is deployed, since this can help the proposed system to adapt to a particular person.
Batch processing is not appropriate for this purpose as it requires offline processing. 
Therefore, we also introduce a Recursive Least Square (RLS) estimation framework that can calibrate the static parameters for a given user.

We can represent the step length in Equation~\ref{eq:distance} as a linear equation with the input features as:
\begin{equation} \label{eq:hn}
{{\bf{h}}[n]} =\begin{bmatrix}
sin(\alpha_{f}[n]) - sin(\alpha_{b}[n])\\ 
sin(\alpha_{f}[n] - \beta_{f}[n]) + sin(\beta_{b}[n] - \alpha_{b}[n])\\ 
1
\end{bmatrix},
\end{equation}
where $n$ is the time index. Similarly, the model coefficients 
become the static parameters we aim to estimate:
%
\begin{equation}
\label{eq:weights}
{{\bf{w}}} = \begin{bmatrix} 
l_1 & l_2 & d_5
\end{bmatrix}^{T}
\end{equation}

When the step length model needs to be updated, e.g., for a new user, 
the system operates in a calibration mode. 
In this mode, the user is asked to walk on a line with a constant step length. During this time, the RLS filter estimates the step length as a product of 
$\bf{h}[n]$ and $\bf{w}$, which are given Equation~\ref{eq:weights} and Equation~\ref{eq:hn}, respectively. 
This estimate is subtracted from the reference step length to find the modeling error. Then, this error is then used to update the parameter estimates $\bf{w}$ using an RLS technique with stability guarantees~\cite{fortescue1981implementation}.

\begin{figure*}[t]
    \centering
    \includegraphics[width=1\linewidth]{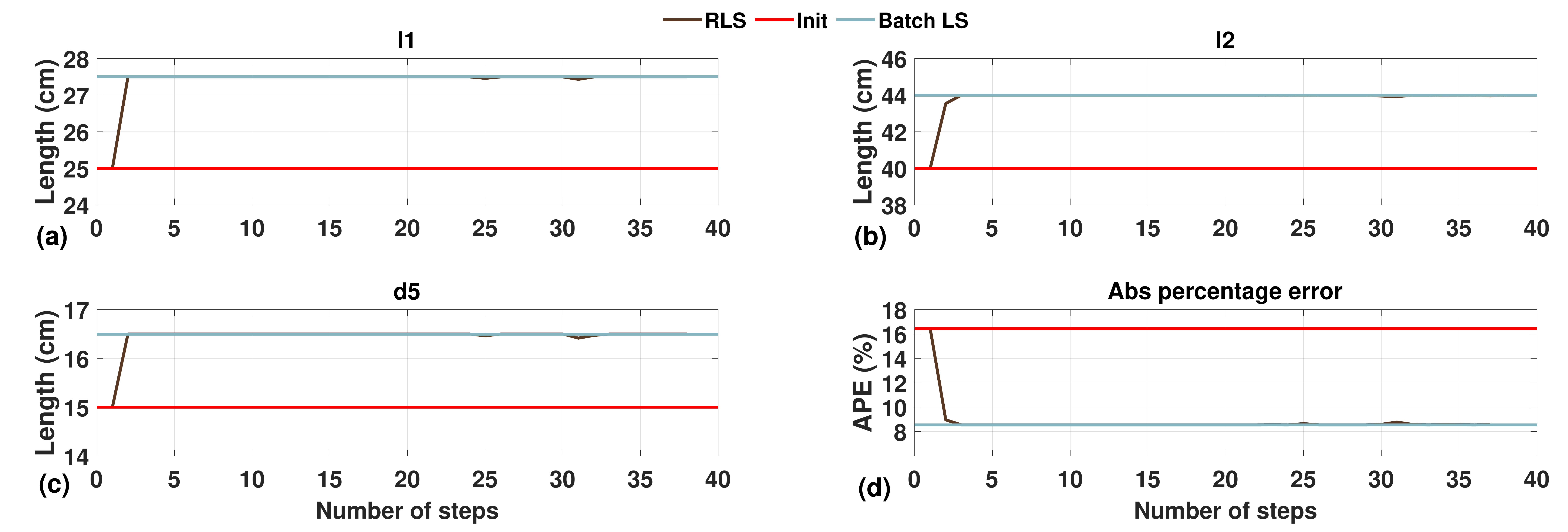}
    \vspace{-2mm}
    \caption{Visualization of how (a) l1, (b) l2, and (c) d5 online updating and converging to the batch LS result and the (d) absolute percentage error of step length estimation using RLS for one subject.}
    \label{fig:rlsresult}
\end{figure*}

Experimental results show that the RLS method improves the step length estimation accuracy by about 3\% on average.
It is effective in reducing the measurement error online, especially for subjects with a high initial error rate. 
For instance, S2 and S5 have 16.44\% and 16.01\% initial MAPE, respectively. Applying RLS estimation, the MAPE are reduced by 7.54\% and 8.05\%, respectively. Figure~\ref{fig:rlsresult} shows how static parameters and the error converge online, as steps increases. The RLS model converges to its final values within five steps. The accuracy of RLS converges also to the batch LS when it is applied to all subjects.
This shows that the RLS approach is effective for estimating the static parameters at runtime.

\noindent \textbf{Offline vs. Online Calibration:}  
The online calibration (Section~\ref{sec:est_stat_param_rls}) can expedite the MGait framework to adjust to new users, but it requires the user to walk in a straight line with a constant step length. 
However, patients with movement disabilities may be unable to walk in a straight line.
In this case, the offline calibration (Section~\ref{sec:est_stat_param_batch}) may be preferred, as it does not require the user to walk in a straight line with a constant step length. 
The user will only walk a few trials without any restriction to do the offline calibration.

\begin{table}[h]
\centering
\caption{Error in step length estimation using offline and online method (without angle correction). RLS represents recursive least square, MAPE represents mean absolute percentage error, and RMSE means root mean square error.}
\label{tab:errorofflinevsonline}
\begin{tabular}{lrrrrrr}
\toprule
     & \multicolumn{2}{c}{Initial}
     & \multicolumn{2}{c}{After Batch LS}
     & \multicolumn{2}{c}{After RLS} \\ 
     \cmidrule(lr){2-3}
     \cmidrule(lr){4-5}
     \cmidrule(lr){6-7}
     & \multicolumn{1}{c}{\begin{tabular}[c]{@{}c@{}}MAPE \\(\%)\end{tabular}} & \multicolumn{1}{c}{\begin{tabular}[c]{@{}c@{}}RMSE \\(cm)\end{tabular}} & \multicolumn{1}{c}{\begin{tabular}[c]{@{}c@{}}MAPE \\(\%)\end{tabular}} & \multicolumn{1}{c}{\begin{tabular}[c]{@{}c@{}}RMSE \\(cm)\end{tabular}} & \multicolumn{1}{c}{\begin{tabular}[c]{@{}c@{}}MAPE \\(\%)\end{tabular}} & \multicolumn{1}{c}{\begin{tabular}[c]{@{}c@{}}RMSE \\(cm)\end{tabular}} \\
     \cmidrule(lr){2-3}
     \cmidrule(lr){4-5}
     \cmidrule(lr){6-7}
S1 &6.38   & 4.03  &5.80 &	3.65 &6.28&	3.99 \\
S2  &16.44 & 12.15 &8.55 &	6.34 &8.90&	6.60 \\
S3 &7.17   & 4.53  &6.88 &	4.31 &6.93&	4.38 \\
S4 &7.65   & 4.99  &5.89 &	3.75 &5.95&	3.79 \\
S5 &16.01  & 8.03  &7.94 &	3.99 &7.96&	3.99 \\
S6 &8.70   & 5.25  &8.07 &	4.86 &8.66&	5.22 \\
S7 &7.88   & 4.47  &7.69 &	4.35 &7.66&	4.35 \\
Avg. & 10.03 &6.21 &	7.26&4.46 &	7.47 &4.62\\\bottomrule
\end{tabular}
\end{table}

\subsection{Analysis of Error Distribution of Angle Measurements}
\label{sec:fit2}
The angle measurements are also subject to error, due to the noisy nature of the sensors. Furthermore, sensors may experience a systematic bias, due to their positioning.
This difference should be accounted for in the models to ensure an accurate estimation of the step length. 
To compensate for these errors, we solve the nonlinear regression problem given in Equation~\ref{eq:opt1} again, by using the static parameters found in Section~\ref{sec:est_stat_param}.
This time, we let the knee and hip angles in Equation~\ref{eq:distance} become free variables. 
That is, the estimation for the $i^{\mathrm{th}}$ step is changed as 
$D_i (\alpha_f, \alpha_b, \beta_f, \beta_b)$ in the problem formulation in Equation~\ref{eq:opt1}.
Similarly, these parameters are constrained within 10\% of their nominal values given by the average of our observations, following the same methodology as in Section~\ref{sec:est_stat_param}.

\begin{figure}[h]
    \centering
    \includegraphics[width=0.7\linewidth]{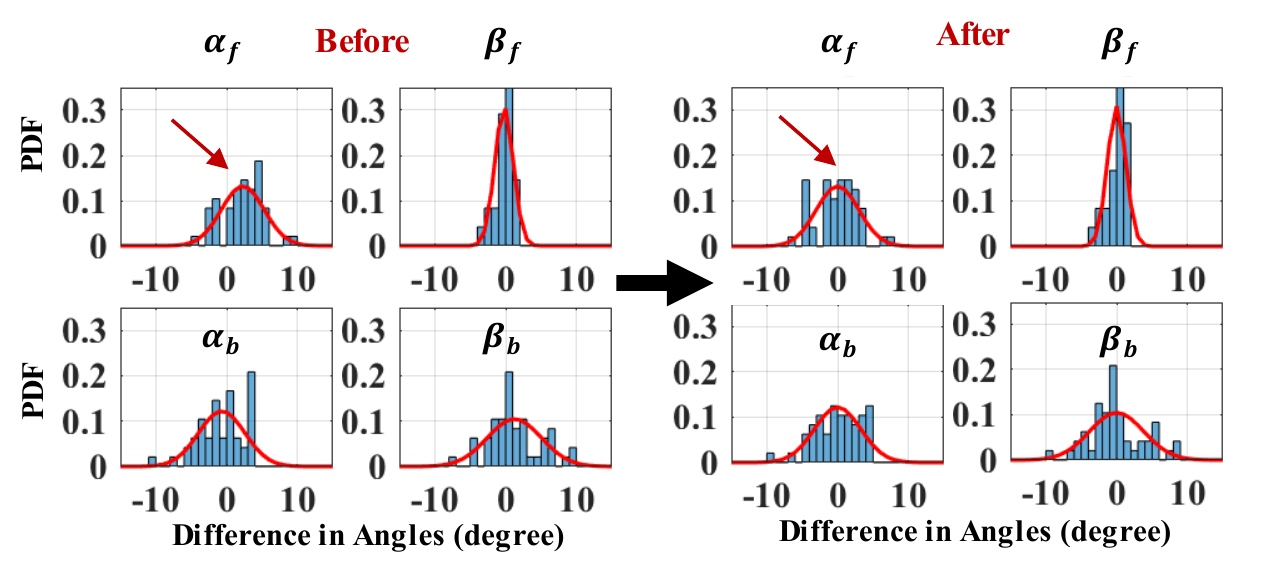}
    \caption{Illustration of the error distribution after removing sensor bias.}
    \vspace{-4mm}
    \label{fig:err_corr_all_before_after}
\end{figure}

The output of the nonlinear regression gives the average knee and hip angles that provide the minimum estimation error. 
Hence, we use the difference between these values and our observations
as the measurement bias. 
For instance, the nonlinear regression finds the hip angle 
of the forward leg at the initial contact point as 
$\alpha_f$~=~24.8\textdegree~for Subject 5. 
The same value is 22.5\textdegree~in our observation dataset.
Consequently, we compute the bias as -2.3\textdegree.
We rectify the sensor bias, by subtracting these empirically determined
values from the knee and hip angles found in real-time. 
After correcting the bias, the error between the best fit and our measured values become zero-mean, as demonstrated in Figure~\ref{fig:err_corr_all_before_after}.

\subsection{Accuracy Evaluation}

\begin{figure}[h]
    \centering
    \includegraphics[width=0.8\linewidth]{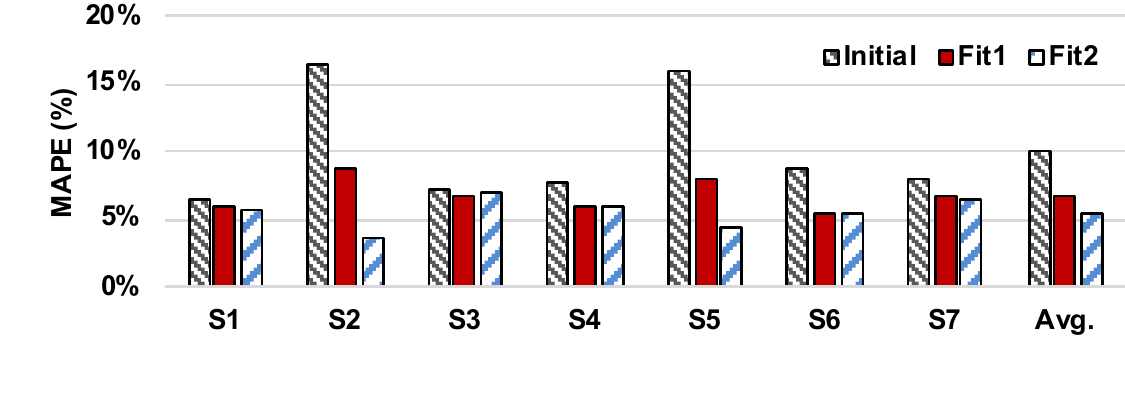}
    \caption{Mean absolute percentage error in step length estimates}
    \label{fig:res_mape}
\end{figure}

We use 70\% of the steps collected from a subject for regression (Sections \ref{sec:est_stat_param} and \ref{sec:fit2}) and the remaining 30\% for evaluating accuracy. The errors reported in this section use the \textit{corrected} static parameters and angle distributions. The average MAPE in step length estimation is about 10.03\%, \textit{when we use the user measurements without any error correction}. 
The error is reduced to about 7.26\% with the fitting of user-specific parameters, as shown in Figure~\ref{fig:res_mape}. 
It is reduced further to about 5.49\%, when we remove the offset in angle measurements.
In particular, the MAPE for Subjects 2 and 5 drops by more than 10\%, since they had a higher initial error due to inaccurate limb measurements. 

We then evaluate the accuracy for step length, stride length and gait velocity for each subject. 
Each row in Table~\ref{tab:mergedtable} corresponds to a different subject, while the columns show the MAPE and RMSE in step length, stride length, and gait velocity, respectively. 
The MAPE and RMSE for step length range from 3.63\% to 6.91\% and 2.83~cm to 5.48~cm, respectively.
The error rates are lower for stride length and gait velocity estimations.  Specifically, the maximum MAPE in stride length and gait velocity is 
6.40\% and 3.59\%, respectively. 
We note that the RMSE value for stride length seems larger, 
since each stride consists of two steps. 
In summary, the results show that the proposed model is able to accurately estimate gait parameters with lower power consumption overheads.

We also compare the proposed approach to simpler techniques, such as integrating the acceleration twice while the feet are in motion~\cite{hannink2017benchmarking}. Even after passing the raw data through a low-pass filter, double integration of acceleration leads to a MAPE of 13.5\%, significantly higher than MGait.

\begin{table}[h]
\centering
\caption{Error in step length, stride length, and velocity with angle correction. The results in this table are for 30\% test data. MAPE represents mean absolute percentage error, and RMSE means root mean square error.}
\label{tab:mergedtable}
\begin{tabular}{lcccccc}
\toprule
     & \multicolumn{2}{c}{Step Length}
     & \multicolumn{2}{c}{Stride Length}
     & \multicolumn{2}{c}{Velocity} \\ 
     \cmidrule(lr){2-3}
     \cmidrule(lr){4-5}
     \cmidrule(lr){6-7}
     & \multicolumn{1}{c}{\begin{tabular}[c]{@{}c@{}}MAPE \\(\%)\end{tabular}} & \multicolumn{1}{c}{\begin{tabular}[c]{@{}c@{}}RMSE \\(cm)\end{tabular}} & \multicolumn{1}{c}{\begin{tabular}[c]{@{}c@{}}MAPE \\(\%)\end{tabular}} & \multicolumn{1}{c}{\begin{tabular}[c]{@{}c@{}}RMSE \\(cm)\end{tabular}} & \multicolumn{1}{c}{\begin{tabular}[c]{@{}c@{}}MAPE \\(\%)\end{tabular}} & \multicolumn{1}{c}{\begin{tabular}[c]{@{}c@{}}RMSE \\(m/s)\end{tabular}} \\
     \cmidrule(lr){2-3}
     \cmidrule(lr){4-5}
     \cmidrule(lr){6-7}
S1   & 5.76 & 4.53 & 4.48 & 6.55 & 3.33 & 0.04 \\
S2   & 3.63 & 3.35 & 3.93 & 7.56 & 0.77 & 0.01 \\
S3   & 6.91 & 5.48 & 3.89 & 6.54 & 1.64 & 0.02 \\
S4   & 5.94 & 4.51 & 3.71 & 6.48 & 2.06 & 0.03 \\
S5   & 4.38 & 2.83 & 1.99 & 2.66 & 0.94 & 0.01 \\
S6   & 5.31 & 3.61 & 4.79 & 6.61 & 2.73 & 0.05 \\
S7   & 6.51 & 4.24 & 6.40 & 7.72 & 3.59 & 0.06 \\
Avg. & 5.49 & 4.08 & 4.17 & 6.30 & 2.15 & 0.03 \\ \bottomrule 
\end{tabular}
\end{table}

\subsection{Evaluation of Real-Time User Feedback}
User feedback on gait quality plays a critical role in patient rehabilitation and health monitoring. 
For instance, feedback on unequal step lengths is an important part of rehabilitation of patients with a leg injury. 
Similarly, gait speed is an important indicator in movement disorders.

\begin{figure}[h]
    \centering
    \includegraphics[width=0.80\linewidth]{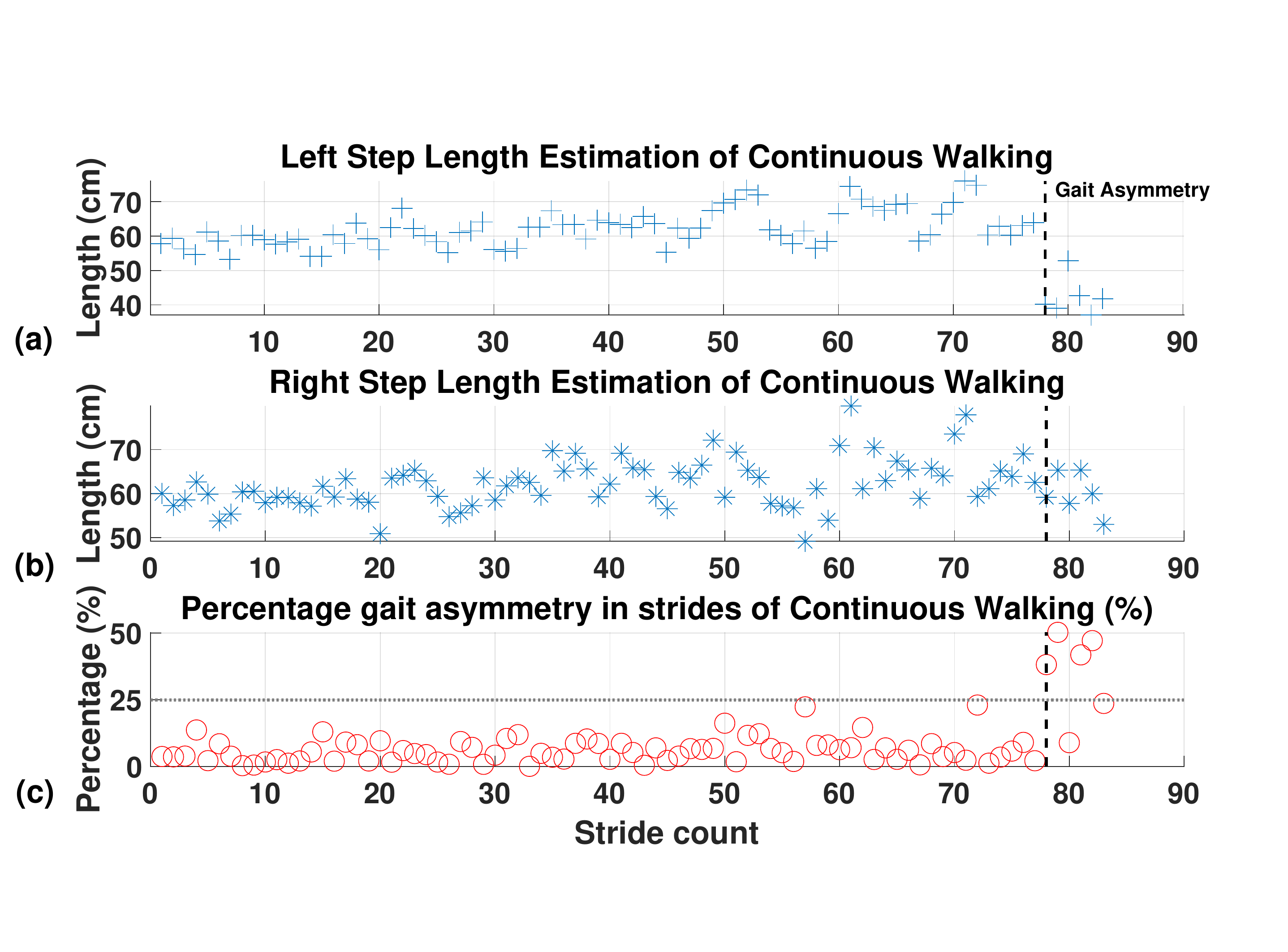}
    \caption{Gait asymmetry detection. (a) shows the left step length estimation, (b) shows the right step length estimation, and (c) shows the percentage gait asymmetry in strides. The grey dot line shows the threshold.}
    \label{fig:feedback_asym}
\end{figure}

An example of the user feedback for one subject given by \textit{MGait} is shown in
Figure~\ref{fig:feedback_asym}. Figure~\ref{fig:feedback_asym}(a) shows the left step length estimation, (b) shows the right step length estimation, and (c) shows the absolute percentage error of the left-to-right difference in strides of continuous walking. \textit{MGait} provides user feedback as soon as the step lengths are asymmetric and the percentage gait asymmetry in a stride exceeds the threshold of 25\%, as shown in  Figure~\ref{fig:feedback_asym}. The subject starts walking, aiming for walking with symmetric steps in strides. After 78 strides, the subject starts making one regular step, followed by a short step (Gait asymmetry).
\textit{MGait} keeps track of the percentage gait asymmetry in a stride. 
Since there are no abnormalities in the first 150 steps, \textit{MGait} does not produce any user feedback. The percentage gait asymmetry in the stride is larger than the threshold of 25\% at $78_{th}$ strides. After this point, \textit{MGait} waits for five steps and compares the average step length before and after the change in variance. Since the percentage gait asymmetry in the stride exceeds 25\%, \textit{MGait} raises feedback. Specifically, the user is notified using a buzzer and an LED.

We evaluated all seven subjects using this feedback algorithm. Overall, 25 strides out of 403 strides are obtained while the user was walking with gait asymmetry. The precision, recall, and f1 score of predicting limping are 1, 0.73, and 0.85, respectively.


\subsection{Sensors Sustainability and Power Analysis}
The reliability of the sensors is crucial for \textit{MGait}'s practicality. In this subsection, we evaluate both sensors' sustainability under continuous monitoring.

\subsubsection{Sensor drift}

\begin{figure}[h]
    \centering
    \includegraphics[width=0.85\linewidth]{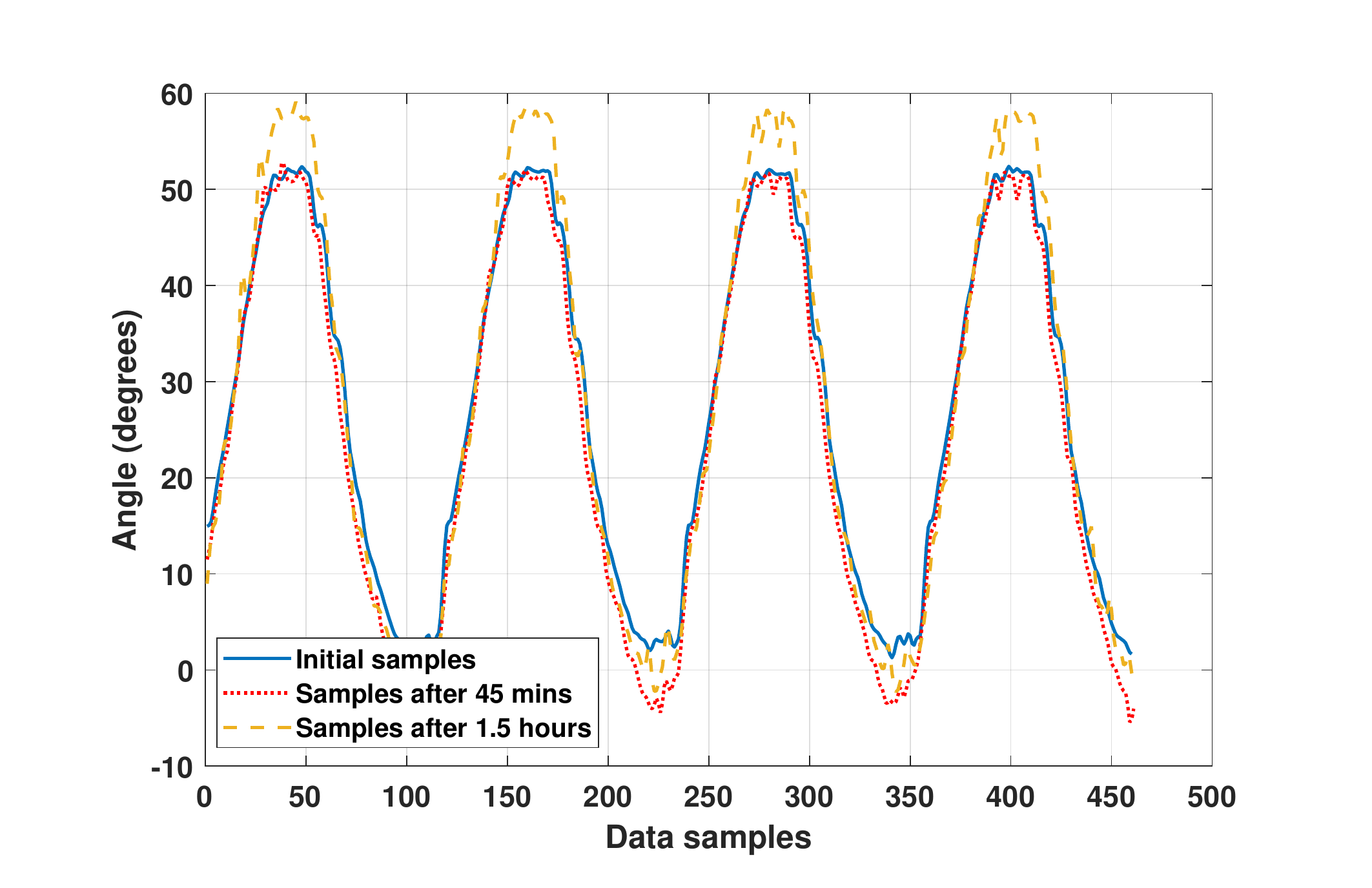}
    \caption{Bending sensors reading under continuous running condition}
    \label{fig:drifting}
\end{figure}

This study is the first work introducing wearable bend sensors to the gait monitoring system. The readers might be interested in the sensing drift of the bend sensor since it is a common problem for all sensors. To this end, a systematic evaluation is performed using a robotic motor joint. We employ a robotic joint performing a 60$\degree$ sinusoidal movement at 1Hz, attach the bending sensor to the joint, and record the sensing data. The experiment lasts 1.5 hours, a time interval that is enough to cover most of the single clinical therapy session. Figure~\ref{fig:drifting} shows that the range of the sensor is initially from 1 to 52$\degree$. After 45 minutes, the sensor shows a sensing range from -2 to 53$\degree$. At the end of the experiments, the sensor exhibits a sensing range from -2 to 59$\degree$. The result indicates that the sensor is reasonably stable in the 1.5 hours of experiments. The sensor drift might be from the friction or the sensors' poor attachment with the joint as time goes on. Also, sports sleeves may slip down after long periods of usage. Therefore, we suggest users align the sports sleeve and calibrate the sensor before each trial to get better performance.

\begin{figure}[h]
    \centering
    \includegraphics[width=0.85\linewidth]{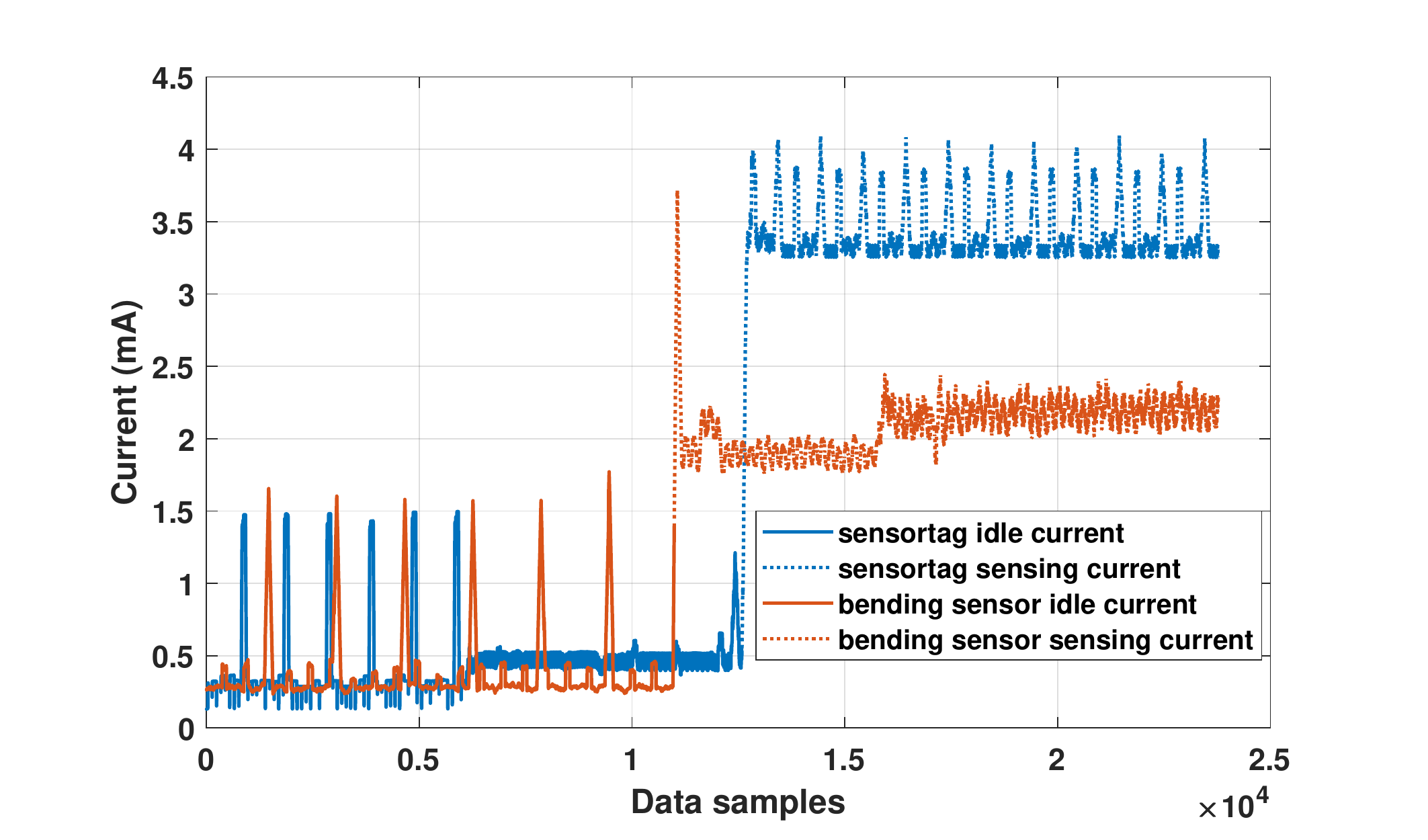}
    \caption{Power consumption of bending sensor and sensortag}
    \label{fig:power}
\end{figure}

\subsubsection{Power evaluation}
Low-power operation is crucial for \textit{MGait} to offer a long-term gait monitoring solution~\cite{park2017flexible}. To explore the power consumption of \textit{MGait}, we measure the actual current consumption of the bending sensors and the IMUs. First, we power the bending sensor with a 150 mAh @ 3.3 V battery and connect a 1 $\ohm$ sensing resistor in series to it. Then we turn the sensor on and start sensing. We record the voltage on the resistor and calculate the current drawn by the sensor. Since the resistor's value is low, the drawn current is essentially the sensor's current consumption. We follow the same steps to measure the IMU's current consumption. Figure~\ref{fig:power} shows the power consumption results. We observe that, for the bending sensor, the average of initial and idle current~(the solid line) consumption is 0.5 mA and the sensing current~(the dashed line) is about 2.3 mA. The idle current consumption of the IMU is 0.5 mA, while  its sensing current is about 3.5 mA. Therefore, when connected to a 150 mAh lithium-ion battery, the bending sensor can run more than 12 days in idle mode, 65 hours in the sensing mode. Similarly, the IMU can run more than 12 days in the idle mode and 43 hours in sensing mode. Note that the power of the bending sensor~(7.6 mW) and the IMU~(11.6 mW) reported here are higher than the values reported in the related work (Table 1). To provide a fair comparison, Table 1 reports only the sensor power consumption~(i.e., processing and wireless communication power consumption are not included) for every work since many of them do not report these power figures. In our case, we include BLE power consumption for bending sensors and the total power consumption for the IMU to evaluate the actual power consumption of \textit{MGait}. In short, \textit{MGait} offers a promising low-power solution for long-term gait monitoring.

\section{Synthetic Dataset Generation} \label{sec:synthetic_dataset_generation}

\subsection{Motivation}

Experimental data collection and labeling is both time- and labor-consuming due to three barriers. 
First, a reliable multi-sensor hardware infrastructure is needed to perform the experiments. Besides calibrating and operating individual sensors, 
the infrastructure should synchronize data from multiple sensors (in our case accelerometer, gyroscope, and bend sensor).
Second, experiments with human subjects require comprehensive experiments and recruitment protocols approved by an institutional review board (IRB). 
Both preparation of protocols and their approval require significant amount of time (a few months). 
Last but not least, representative datasets require multiple volunteers with diverse attributes, such as age, gender, and height.

Throughout this study, sensor data and step length data were collected for more than 800 steps from seven subjects following an IRB protocol. Although this data is sufficient for a thorough experimental evaluation, additional data points may catalyze future research. 
%
The synthetic data generation can augment experimental data to create a richer dataset. Synthetic data has been previously used to make significant advances in many domains. For instance, in the Network-on-Chips~(NoC) domain, real-world workloads available at design time do not represent all the traffic patterns that will be encountered by the network in the future~\cite{soteriou2006statistical}. Furthermore, running multiple workloads to test the network under varying traffic conditions is time-consuming. In order to address this, a number of synthetic workloads were developed for NoCs~\cite{ogras2013modeling,mandal2019analytical}. These synthetic traffic patterns speed up the design and test of NoCs architectures, leading to significant advances in the field.

To this end, we present a technique to generate synthetic training and test data with the help of our existing experimental dataset to enable gait analysis research with more data. To the best of our knowledge, this is the first dataset that provides two sensor modalities for gait analysis. The synthetic data, along with the original sensor data, will aid future researchers to develop novel methods of gait analysis while avoiding the tedious step of data collection. 

\subsection{Background}
Multiple data generation methods using deep neural networks have been proposed in recent years. Goodfellow \textit{et al.}~\cite{goodfellow2014generative} presented Generative Adversarial Networks (GANs) in 2014 for the first time. GANs have been implemented for a variety number of applications in image, text, and sensor dataset generation. Alzantot \textit{et al.}~\cite{alzantot2017sensegen} proposed an architecture for generating wearable sensor data, including accelerometer and gyroscope values using Gaussian Mixture Density Network connected to a recurrent neural network  (RNN). 
Although GAN models can generate new random plausible examples for a given dataset, they do not have labels for the data, which limits their effectiveness for supervised learning tasks. To solve this problem, Mirza \textit{et al.}~\cite{mirza2014conditional} proposed a Conditional Generative Adversarial Networks (CGAN). It includes the label information in the input of the discriminator and the generator as an additional input layer.

\subsection{Proposed Technique}

We implement a CGAN to generate the step length dataset. The objective function of CGAN is shown below:
\begin{equation*}
\label{eq: GAN}
\min _{G} \max _{D} \mathbb {E}_{x \sim p_{\text {data}}}\left [{\log D\left ({x}\right|{y})}\right ] +\mathbb {E}_{z \sim p_{z}} \left [{\log \left ({1-D\left ({G\left ({z}\right|{y} )}\right )}\right )}\right ].\tag{1}
\end{equation*} 
The idea behind GAN is a two-player minimax game. GANs consist of both generator and discriminator models. The generator is used to generate new domain samples from latent space, while the discriminator classifies whether the input samples are real or generated. 
Ideally, in the end, the generator can generate samples that are hard for the discriminator to distinguish whether they are real or synthetic. Similarly, the discriminator is working decently when it is able to tell whether samples are from the real distribution or generated.

In the implementation of CGAN, we employ an embedding layer for the labels, then concatenate it with our input data samples as the inputs. Therefore, the model knows which class the current data belongs to (for the discriminator) and which class the generated data should be labeled as (for the generator). In such a way, labels, which are the additional information are attached to the input samples. The improvement shows more stable and faster training, and most importantly, the generated samples have labels. 

\begin{figure}[h]
    \centering
    \includegraphics[width=1\linewidth]{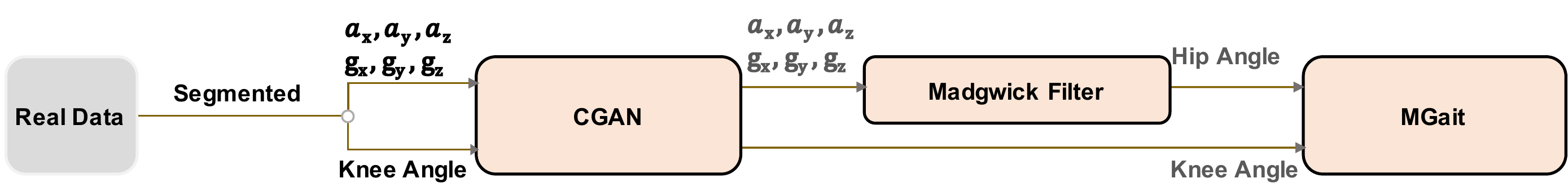}
    \caption{Overview of the synthetic data generation flow}
    \label{fig:ganflow}
\end{figure}

Figure~\ref{fig:ganflow} describes the flow of generating a synthetic dataset using CGAN. Algorithm~\ref{algo:CGAN} shows the pseudo-code for how we train the CGAN. From the real data, we segmented windows based on each walking step. In a specific window, there are 3-axis accelerations, 3-axis angular velocity, and knee angle data. We then gave these data to the CGAN as the input. In our case, the dimension of data is 7 (3+3+1) and the data length is 41. Each window is, on average, 1.64 seconds in duration, since the sampling rate is 25~Hz. Note that not every window is exactly 1.64~seconds. For the shorter width windows, we pad the sensor data with zeros. We also embedded the step length as an additional layer to the data. The CGAN takes these inputs to train the model.
As mentioned above, in each epoch, we first use real data from the input and the noise from latent space to train the discriminator to differentiate the real data from the synthetic data. Secondly, we use the generator to generate the synthetic data and use this synthetic data to train the whole CGAN model, while disabling the training of the discriminator. In other words, we only train the generator here. After enough epochs, as both the loss of both discriminator and generator converges, CGAN is well-trained. We then use the generator to generate the synthetic data with labels. To get the hip angle, we need to use the Madgwick filter to process the synthetic raw IMU data again. Finally, we use the \textit{MGait} model to estimate the step length from the synthetic data and compare the accuracy.

\begin{algorithm}[H]
\footnotesize
\SetAlgoLined
\label{algo:CGAN}
\KwIn{a batch of k samples $\{e^{(1)}_{real},\dots,e^{(k)}_{real}\}$, where $e^{(k)} = (x^{(k)}_{real},y^{(k)}_{real})$. $x^{(k)}_{real}$ here is the 6-axis IMU data and the knee angle, $y^{(k)}_{real}$ here stands for the label, which is the step length for this window.}
\KwOut{a batch of k synthetic samples $\{e^{(1)}_{syn},\dots,e^{(k)}_{syn}\}$. where $e^{(k)} = (x^{(k)}_{syn},y^{(k)}_{syn})$, $x^{(k)}_{syn}$ here is the 6-axis IMU data and the knee angle, $y^{(k)}_{syn}$ here stands for the label, which is the step length for this window.}
\For{numbers of training iterations}{\

\begin{itemize}
\item Sample a minibatch of $s$ noise samples \\$\{e^{(1)}_{syn},\dots,e^{(s)}_{syn}\}$ from noise prior $p_{g}(z)$.
\item Sample a minibatch of $s$ samples
$\{e^{(1)}_{real},\dots,e^{(s)}_{real}\}$ from data distribution $p_{data}(e)$.
\item Update the discriminator by increasing its stochastic gradient using Adam optimizer.
\\
\item Sample a minibatch of $s$ noise samples \\$\{e^{(1)}_{syn},\dots,e^{(s)}_{syn}\}$ from noise prior $p_{g}(z)$.
\item Update the generator by decreasing its stochastic gradient using Adam optimizer.

\end{itemize}
}
\caption{Training of the CGAN}
\end{algorithm}

\subsection{Visualization and Evaluation of the Synthetic Data}

\begin{figure*}[t]
    \centering
    \includegraphics[width=0.90\linewidth]{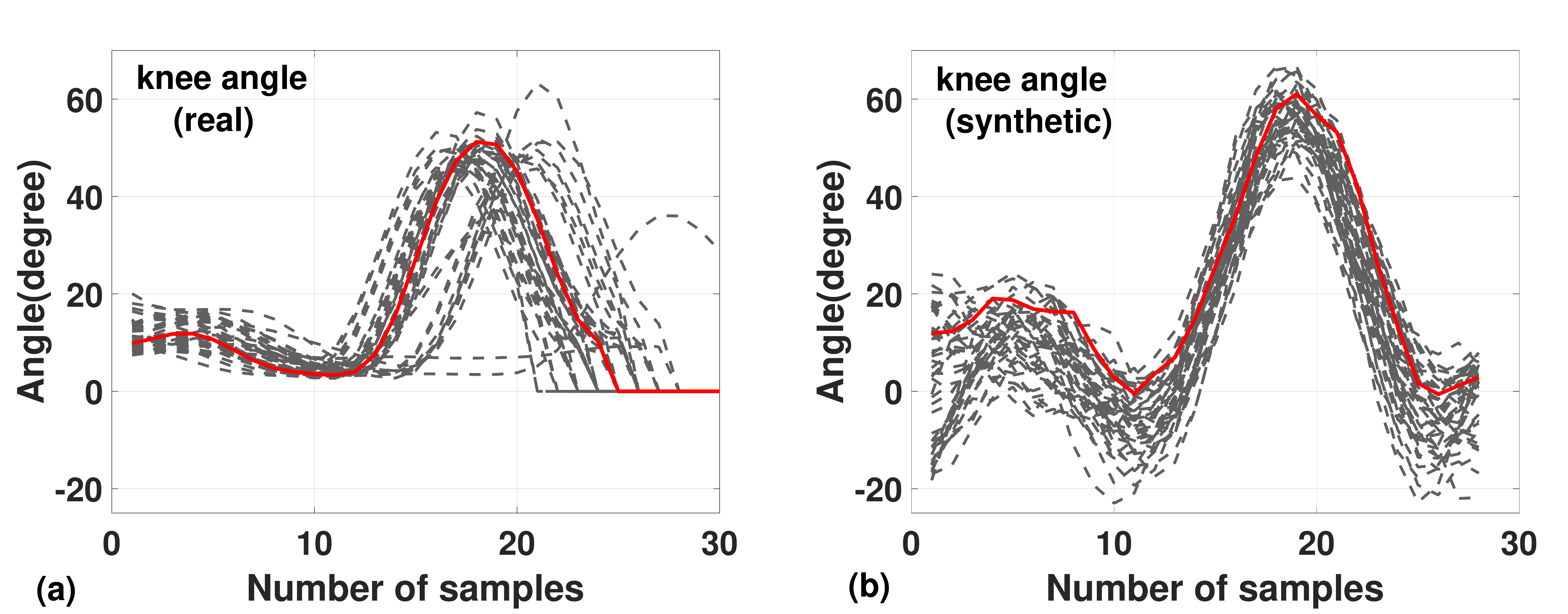}
    \includegraphics[width=0.90\linewidth]{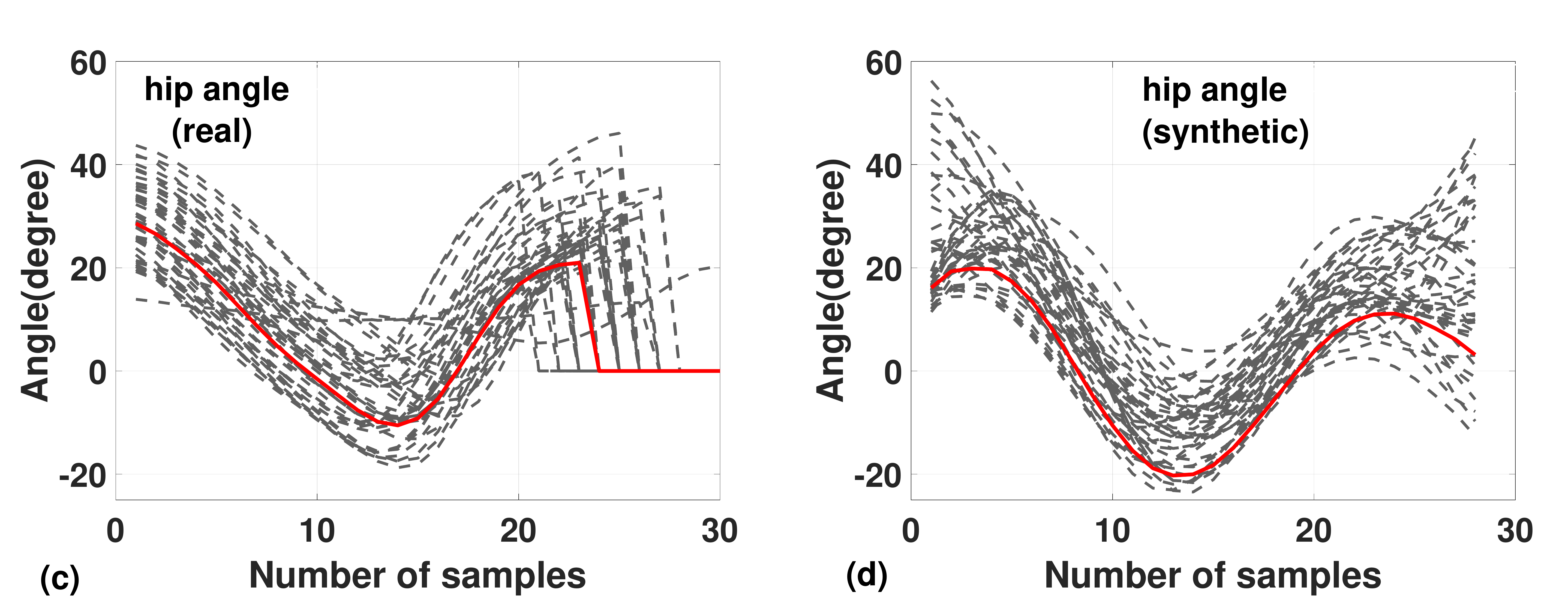}
    \vspace{-1mm}
    \caption{One window of (a) \textit{measured knee} angle, (b) \textit{synthetic knee} angle, (c) \textit{real hip} angle, and (d) \textit{synthetic hip} angle for a given step length. The grey dashed lines show different instances of the data, while the red line shows a \textit{representative} window.}
    \label{fig:ganhip}
\end{figure*}

Figure~\ref{fig:ganhip}(a) plots the knee angle based on our experimental data. The angles for each user are overlayed on top of each other to show the pattern and variations. 
Figure~\ref{fig:ganhip}(b) shows that the synthetically generated data leads to a very similar pattern with the real knee angle. 
Similarly, we observe a good match between the hip angle based on the experimental data and synthetic data, as shown in Figure~\ref{fig:ganhip}(c) and Figure~\ref{fig:ganhip}(d).
We also note that the synthetic raw data also has a similar pattern with the real raw data. 
We do not plot all 6-axis IMU data here for the readability. 

The quality of the synthetic data is evaluated first using statistical metrics, including mean, standard deviation, and correlation coefficients of both measured and synthetic data over a window of samples that correspond to a single step. 
The mean value of the synthetic data is within 1$^{\circ}$ degree of the measured data for both knee and hip angles, as shown in Table~\ref{tab:metricssyntheticdata}. 
The synthetic data has a slightly higher standard deviation than the real data since its range increases when the GANs try to capture all data patterns from the real data.
Table~\ref{tab:metricssyntheticdata} also compares the pairwise correlation coefficients of all data windows. 
The mean correlation coefficients of measured and synthetic data are 0.71 and 0.79, respectively, for the hip angle. 
Similarly, the mean correlation coefficients of the measured and synthetic knee angles are 0.77 and 0.90. The real data has a lower correlation coefficient since it contains some outliers from sensors reading, resulting in a low correlation coefficient. We fine-tune the GANs such that the generated synthetic data will not learn the outlier data patterns, which leads to a higher correlation coefficient in generated data.

In addition to the statistical analysis, we evaluate the quality of the synthetic data using a ``representative'' window of hip and knee angles, as shown in Figure~\ref{fig:ganhip}. 
Each window of data corresponds to a different step in our data set. 
To find the representative window, we first compute the Euclidean distance between each window in the data set. A large distance between a given window and others imply that the corresponding window is more likely to be an outlier. Therefore, the window with the smallest average Euclidian distance to other windows is chosen as the \textit{representative window} for the corresponding angles, as illustrated with red lines in Figure~\ref{fig:ganhip}. 
We observe that the representative instances of synthetic and real data are very similar in terms of data pattern and the values for both hip and knee angles, as shown in Figure~\ref{fig:ganhip}(a), (b), (c), and (d), respectively.

To show that our synthetic data can be used for the gait analysis, we also compare step length estimations using real data and the synthetic data in Table~\ref{tab:syntheticdata}. Without doing any regression, the 6400 steps synthetic data achieve a 9.39\% MAPE in step length estimation, while the real data is 10.03\%. After the batch LS and RLS process, the MAPE of synthetic data reduced to 8.17\%,  and it remains the same for the RLS approach. For step estimation, using both initial and regressed synthetic data give similar MAPE~(within 1\%) when compared to the real data.

We also design the experiments using synthetic data for training. First, a set of synthetic data is generated. Then, we set three configurations to train the batch LS and RLS models: 
\begin{enumerate}
\item {Use only real data,} 
\item {Use only synthetic data,} 
\item {Use equal amounts of real and synthetic data.}
\end{enumerate}
Finally, we test the model with only real data. As shown in the first, third, and fourth rows of Table~\ref{tab:syntheticdata}, adding synthetic data in training gives a similar MAPE compared to training with only the real data. Therefore, the synthetic data has a similar pattern to the real data, and it can be used for step length estimation when real user data is not available or limited.

\begin{table}[t]
\centering
\caption{Statistical metrics of real data and synthetic data. std represents standard deviation.}
\label{tab:metricssyntheticdata}
\setlength{\tabcolsep}{2.5pt} 
\begin{tabular}{lrrrrrr}
\toprule
     & \multicolumn{3}{c}{Knee Angle}
     & \multicolumn{3}{c}{Hip Angle} \\ 
     \cmidrule(lr){2-4}
     \cmidrule(lr){5-7}
     & \multicolumn{1}{c}{\begin{tabular}[c]{@{}c@{}}Mean\end{tabular}} & \multicolumn{1}{c}{\begin{tabular}[c]{@{}c@{}}Std.\end{tabular}} & \multicolumn{1}{c}{\begin{tabular}[c]{@{}c@{}}Correlation Coefficients\end{tabular}} & \multicolumn{1}{c}{\begin{tabular}[c]{@{}c@{}}Mean\end{tabular}} & \multicolumn{1}{c}{\begin{tabular}[c]{@{}c@{}}Std.\end{tabular}} & \multicolumn{1}{c}{\begin{tabular}[c]{@{}c@{}}Correlation Coefficients\end{tabular}} \\
     \cmidrule(lr){2-4}
     \cmidrule(lr){5-7}
Real data &14.45 &	15.05 &0.77& 9.76 &12.56 &	0.71\\\midrule
Synthetic data &14.59 &	21.11 &0.90& 9.33 &14.14 &	0.79 \\ \bottomrule

\end{tabular}
\end{table}

\begin{table}[t]
\centering
\caption{Error in step length estimation using real and synthetic data (w/o angle correction). RLS: recursive least square, MAPE: mean absolute percentage error, and RMSE: root mean square error.}
\label{tab:syntheticdata}
\begin{tabular}{llrrrrrr}
\toprule
     \multicolumn{1}{l}{Train with}
     & \multicolumn{1}{l}{Test with}
     & \multicolumn{2}{c}{Initial}
     & \multicolumn{2}{c}{Batch LS}
     & \multicolumn{2}{c}{RLS} \\ 
     \cmidrule(lr){3-4}
     \cmidrule(lr){5-6}
     \cmidrule(lr){7-8}
     &
     & \multicolumn{1}{c}{\begin{tabular}[c]{@{}c@{}}MAPE \\(\%)\end{tabular}} & \multicolumn{1}{c}{\begin{tabular}[c]{@{}c@{}}RMSE \\(cm)\end{tabular}} & \multicolumn{1}{c}{\begin{tabular}[c]{@{}c@{}}MAPE \\(\%)\end{tabular}} & \multicolumn{1}{c}{\begin{tabular}[c]{@{}c@{}}RMSE \\(cm)\end{tabular}} & \multicolumn{1}{c}{\begin{tabular}[c]{@{}c@{}}MAPE \\(\%)\end{tabular}} & \multicolumn{1}{c}{\begin{tabular}[c]{@{}c@{}}RMSE \\(cm)\end{tabular}} \\
     \cmidrule(lr){3-4}
     \cmidrule(lr){5-6}
     \cmidrule(lr){7-8}
Real data & Syn. data  & 9.39 &5.63 &	8.17 &4.90 &	8.17 &4.90\\ \midrule
Real data & Real data & 10.03 &6.21 &	7.26&4.46 &	7.48 &4.61\\\midrule
Real + Syn. data & Real data & 9.36 &5.61 &	7.28&4.47 &	7.60 &4.70\\\midrule
Syn. data & Real data  & 9.13 &5.51 &	7.29 &4.47 &	7.70 &4.78\\ \bottomrule
\end{tabular}
\end{table}

\section{Conclusions and Future Work} \label{sec:conclusion} 
Gait analysis plays a crucial role in patient rehabilitation and health monitoring of movement disorder patients. This paper presented \textit{MGait}, a wearable CPS for step length, stride length, and gait velocity estimation. We start with a wearable device that incorporates bend sensors and IMUs to measure the knee and hip angles of the subject.
We then use the angles to determine the gait parameters of the subject at runtime with constant complexity. Experimental evaluations using seven subjects showed that our approach is able to estimate the step length, one of the most widely used gait parameters, with 5.49\% error on average.
We propose a synthetic data generation technique, based on CGAN, and we will give the access of both real data and synthetic data to the public. We plan to use the \textit{MGait} framework for studies with movement disorder patients and enable a robust ecosystem for health monitoring and rehabilitation.

\section{Acknowledgments}

The authors would like to thank Professor Narayanan Krishnamurthi and Niveditha Muthukrishnan at Arizona State University for the assistance in accessing the GAITRite equipment. This research was funded in part by NSF CAREER award CNS-2114499.

\bibliographystyle{acm}
{\footnotesize{\bibliography{references/wearable_iot,references/embedded_refs,references/flexible,references/health_refs,references/gait_refs}}}

\end{document}